\documentstyle[11pt]{article}
\newcommand{\ket}[1]{|#1\rangle}
\newcommand{\braket}[2]{\langle #1|#2\rangle}
\newcommand{\matel}[3]{\langle #1|#2|#3\rangle}
\newcommand{\proj}[2]{|#1\rangle \langle #2|}
\newcommand{\nsq}[1]{\| #1\|^{2}}
\begin{document}
\begin{center}
   {\Large \bf Quantum Theory of Observers}\\
   \vspace{.20in} {\large Jose L. Balduz Jr}\\
   {\large \it Department of Physics}\\
   {\large \it Mercer University, Macon, GA 31207, USA}\\
   \vspace{.20in} {\large August 16, 2001}\\ 
   \vspace{.40in} {\footnotesize \bf Abstract}
\end{center}
\begin{quote} \footnotesize
      This is an attempt to create a consistent and non-trivial
      extension of quantum theory, describing in detail the quantum
      measurement process. A tentative but concrete model is presented,
      based on the concept of multiple observer/participators, represented
      by separate state vectors. The evolution is deterministic, and in the
      chaotic regime implies approximate adherence to the Born rule for
      probabilities. The model is applied in a number of contexts: simple
      detectors, multi-state selectors, and intermittent systems. The results
      are consistent with phenomenology. We also consider more speculative
      applications, including specific spin and position `observables.'
      Finally the outlook for the model is discussed, and its relation to
      other work.
\end{quote}
\vspace{.25in}
      The motivation for this work is the lack of a unified dynamics in
      standard quantum theory, a fixed rule specifying the change of the
      system with time. Instead there are two incompatible rules: the
      Schroedinger Equation(SE) whenever no measurement is taking place,
      and the Born rule, which prescribes that the result of a measurement
      is random but occurs with frequencies given by the Born
      probabilities(BP). In the search for a unified dynamics we assume SE
      must be modified to yield measurement results. Thus we define the
      {\em multiple observer model}(MOM), in the Introduction($\S\S 1,2$)
      and Part I(\S\S3-7): No attempt is made to describe the full history
      of any particular measurement, rather the collapse process itself is
      described by use of MOM. In Part II(\S\S8-9), MOM is applied to more
      realistic measurement processes. In the Conclusion(\S\S10-13) the MOM
      is placed in context. General principles and specific competing points
      of view are covered, and then a classification scheme is presented to
      put the whole in perspective.
\part*{Introduction}
\addcontentsline{toc}{part}{Introduction}
\begin{quotation} \footnotesize
   When in doubt, enlarge the quantum system. Then it is found that the
   division can be so made that moving it further makes very little difference
   to practical predictions. Indeed good taste and discretion, born of
   experience, allows us largely to forget, in most calculations, the
   instruments of observation.\cite{Bell1}

   The continuing dispute about quantum measurement theory is\ldots
   between people who view with different degrees of concern or complacency
   the following fact: so long as the wave packet reduction is an essential
   component, and so long as we do not know exactly when and how it takes
   over from the Schroedinger equation, we do not have an exact and
   unuambiguous formulation of our most fundamental
   physical theory.\cite{Bell2} {\em J. S. Bell}
\end{quotation}
\section{A Phenomenological Model}
      After discussing the measurement situation generally, we introduce 
      the concept of the {\em pointer basis}, and then concentrate on the
      process which SE fails to describe: the step whereby the measurement's
      outcome is decided, the collapse. We introduce a simple model based on
      the dynamics of a random walk in probability space.
      This will serve as the phenomenological basis for MOM.

\subsection{About measurements}
      To motivate the assumptions of this model, let us consider
      how measurements are usually thought to occur. The objects involved
      are divided into two sets, the micro(quantum)-systems and
      the macro(classical)-systems. Microsystems do not obey
      classical laws of physics: They evolve by SE like all quantum systems.
      Classical systems are detectors and state preparation devices: They are
      devoid of quantum ambiguities. The measurement
      process is as follows. A microsystem is prepared in a definite known
      state by use of a classical preparation apparatus. For self-consistency
      we may consider this to be the result of a previous measurement.
      The microsystem is then allowed to evolve for a finite period of time,
      as correlations are created between the
      microsystem and another system. The second system evolves, 
      becoming correlated with another system, and so on \ldots
      Thus a chain of systems is obtained wherein each member's state
      is correlated perfectly with the state of the original
      microsystem: `von Neumann's chain'. 
      At some point, however, there is at least one member of the
      chain which is a classical system. It cannot partake in any
      quantum ambiguity, so it must make a choice among some set of
      classically allowed states. The classical apparatus comes to a definite
      arrangement, and the correlations induce a cascade along the chain
      which leaves all the systems in compatible states. The measurement has
      been performed.
      In the orthodox view, the particular chain of systems used is not
      important, nor the place of transition between microsystems and
      classical apparatus. What counts is that there be a microsystem at one
      end, a classical system at the other end, and the
      appropriate interactions between the chain elements so the correct
      correlations are established. The details of the process do not affect
      the probabilities for the various classically observable results and
      are therefore unphysical, unobservable.

      Experimenters routinely link together such chains of systems in order
      to perform measurements. Nevertheless, there is a theoretical problem.
      In the transition region between quantum and classical behavior there
      must be at least one `hybrid' system: It must interact
      quantum-mechanically in order to become correlated with the previous
      (quantum) system, but it must also make a choice among the various
      observable outcomes.
      This system must obey neither quantum nor classical laws of
      evolution. Its existence would verify the objectivity of the collapse
      process. But then it should be possible to devise an experiment to
      violate the statistical predictions of quantum theory, i.e. BP. 
      If no such a system is ever identified and isolated,
      it could be that the transition from quantum to classical behavior is
      illusory, measurements are not well described by the above picture, and
      there is no collapse, {\em or} that this transition is absolute,
      there being only microsystems and detectors but no `hybrids'. Either
      case is theoretically troublesome.

      Assuming such hybrid systems do exist, what kinds of systems are
      they (i.e. which degrees of freedom are involved), what are
      their dynamics, and what deviations from standard quantum predictions
      should we expect? Here we assume the answer to the first question to be
      given: Certain systems are assumed to be special in this regard.
      With this information we postulate a conceptual model and detailed
      dynamics. From these we calculate what unusual results we should expect
      in ordinary measurements, where the quantum/classical transition does not
      play a major role. They are limited to small deviations from BP.

\subsection{Pointer-basis situation}
      Simple measurements are most easily understood by invoking the 
      existence of a {\em pointer basis}. This is a special basis for the
      small, or `local' Hilbert space. 
      A pure local state $\psi$ can be written as
\[ \ket{\psi} = \sum_{i=1}^{N} s_{i} \ket{i}\;, \]
      and the complete, or `global state' $\Psi$ as
\[ \ket{\Psi} = \sum_{i,a=1}^{N,A} C_{ia} \ket{i}\otimes \ket{a}\;,  \]
      where $\ket{i}$ is a pointer-basis element and $\ket{a}$ is a basis
      element in the Hilbert space of the environment degrees of freedom.
      The pointer-basis assumption is that the self energy of $\psi$ 
      and the interactions between the local system and environment are 
      diagonal in the pointer basis. It must be emphasized, that in case
      the interactions {\em do not}\/ commute with the self energy, 
      there is {\em no} pointer basis: The validity and usefulness of this
      concept varies with the system.
      This is relevant to the task of experimenters. Their purpose is to
      set up a situation, where the system of interest can
      find its way onto one state from a set of states which comprise the
      set of possible experimental results. They will know that one of these 
      states has been reached, and which one, by the fact that some part of
      the experimental apparatus, the {\em pointer(!)}, has reached one of the
      corresponding positions. This is the origin of the term, pointer basis.
      The apparatus self energy must be such, 
      that once the pointer gets to one of these positions, it stays 
      there long enough for the fact to be recorded elsewhere.
      If the self energy quickly drives it away, the moment will pass 
      unnoticed, and will not constitute a practical measurement.
      The reason to introduce the pointer basis in the micro-system is the
      same: Its existence is a {\em prerequisite} for the {\em possibility}
      of any measurement. Below, we will simplify the dynamics even further
      by assuming that all self energies vanish, and that the interactions
      can be written as
\[ \hat{H}=\hat{H}_{int}=\sum_{ia} H_{ia} \proj{ia}{ia}\;,
                 \;\; \ket{ia}\equiv \ket{i}\otimes \ket{a} \;.  \]
      For future use we define a
      {\em global state reduced density matrix} (GDM)
      $\hat{Q}$, which is the global density matrix reduced to a 
      local-Hilbert space density matrix by tracing over the environment:
\begin{equation}
      \hat{Q} = \sum_{a} \matel{a}{\Psi \Psi^{\dagger}}{a} 
              = \sum_{ij} Q_{ij} \proj{i}{j}\;,\;\;\;
               Q_{ij} = \sum_{a} C_{ia} C_{ja}^{\ast}\;.  \label{eq:GDM}
\end{equation}

      These assumptions simplify the dynamics greatly while maintaining the
      essential features of common measurements. The outcome of such
      a measurement is one of the pointer-basis elements. How natural is
      the pointer-basis assumption? On the one hand, all physical systems
      violate it to some extent, since interactions depend on coordinates
      such as particle positions, whereas self-energies contain derivative
      terms(kinetic energy): The two do not commute. On the other hand, 
      usually every effort is made to diminish the effect of the
      kinetic terms so that measurements are stable. Eventually every 
      measurement outcome becomes undone by natural processes,
      but if it lasts long enough to be recorded, this is not a problem.
      The pointer-basis assumption is 
      justified when considering standard measurement procedures.

\subsection{Probability random walk}
      If the system of interest, which is undergoing a measurement, has
      interacted with its environment long enough 
      (i.e. over one or more decoherence times),
      it can be represented by a density matrix which is diagonal in
      the pointer basis:
\[ Q_{ij}\rightarrow Q_{i}\delta_{ij}\;. \]
      The task for a model of measurements is to select one of these basis
      states as the outcome. We can do this in a simple way, by postulating
      that the system undergoes a {\em random walk}. This is the essence of 
      Pearle's `models for reduction', which are defined stochastically with
      no attempt to explain the process in a deeper way \cite[Pearle]{stoc3}.
      This is a useful paradigm, a context for discussion of such models,
      which must add some mechanism, random or deterministic, to generate
      the random walk. Such a mechanism may lead to predictions which are in
      violation of BP, or otherwise contrary to standard quantum theory,
      providing a basis for experimental judgement.
      MOM, described below, {\em has} deviations from BP, while
      the basic Pearle model and the simple model described here are 
      not specific enough to be tested experimentally.

      Let us consider a two-level system first. The density matrix is
      represented by a single real number $z$ between zero and one, which
      according to the Born rule,
      is the probability that one of the two states is the
      measurement outcome. The random walk assumption is that, after each
      time interval $\delta t$, $z$ will change to either $z+\delta z$ or
      $z-\delta z$, where $\delta z$ is constant, with equal probability
      ({\em symmetry condition}). The only exception is that, if $z$ reaches
      0 or 1, the process stops, as the measurement outcome has been decided.
      In the ideal case, $\delta z$ is a constant integer fraction of 1, and 
      the initial value $z$ is an element of the set $\{0,\delta z, 2\delta z,
      \ldots ,1-2\delta z, 1-\delta z, 1\}$. This leads to the confirmation
      of the Born rule: The measurement outcomes occur with
      probabilities $z$ and $1-z$. If either $z$ or $\delta z$ doesn't satisfy
      its condition, the probabilities will deviate from BP by at most
      $\delta z$. The average time before an outcome is reached is finite and
      proportional to $\delta t/(\delta z)^2$. There are some paths which do
      not terminate at either 0 or 1, but these are of measure zero. These
      simple results can be deduced from, for example, Feller\cite{Feller}.
      In the case of a system with N states, the situation is similar but
      slightly trickier. The symmetry condition can still be maintained, but
      we replace $\delta z$ with the average over $N-1$ directions,
      $\overline{\delta z}$.

      Given a small enough $\overline{\delta z}$, 
      this crude model will reproduce BP for measurement outcomes.
      However, it is inadequate as
      a serious physical description of the physical processes involved in
      several ways.
      The system undergoing measurement is chosen arbitrarily.
      Unless this system is the last one in von Neumann's chain,
      the particular choice will affect the oucome probabilities, if
      $\overline{\delta z}$ varies from system to system.
      The choice for pointer basis is arbitrary, whereas it should be dictated
      by some features of the physical systems under consideration.
      The interactions of the system with its environment should determine
      the choice, and whether the pointer-basis assumption applies at all.
      The cause of the random changes in the system is unspecified. If there
      is no explanation for this, the mystery of the state-vector collapse
      is not solved but merely relocated. Is there really such a thing
      as {\em fundamentally} random behavior in the universe?
      The values of $\delta t$ and $\overline{\delta z}$ are arbitrary.
      Thus they may change from system to system, and can take on values 
      such that no experiments will reveal the probability deviations.
      A more detailed model could constrain or predict these numbers, 
      and hence could be refuted or confirmed by experiments.

\section{The Multiple Observer Model}
\begin{figure}[!htbp]
\setlength{\unitlength}{.9cm}
\begin{picture}(18,24)(-1.5,-1)
  \multiput(1,19)(6,0){3}{\dashbox{.5}(4,4){}}
  \multiput(1,11)(6,0){3}{\dashbox{.5}(4,4){}}
  \multiput(2.5,3)(9,0){2}{\dashbox{.5}(4,4){}}
  \put(13.1,19.1){\framebox(3.8,3.8){}}
  \multiput(2.6,3.1)(9,0){2}{\framebox(3.8,3.8){}}
  \put(9,21){\circle{2}}
  \put(2,14){\circle{1.6}}
  \put(4,12){\circle{1.4}}
  \put(8.5,13.5){\circle{2.5}}
  \put(9.5,12.5){\circle{2.5}}
  \put(15,13){\circle{3}}
  \put(14.8,12.8){\circle{.7}}
  \put(4,5){\circle{2}}
  \put(12.5,5){\circle{1.2}}
  \put(14.5,5){\circle{1.2}}
  \put(0,18.5){\makebox(0,0)[tl]
            {\begin{tabular}{ll}
               a) & Set $\cal U$ of classical\\
                  & degrees of freedom.
             \end{tabular}}}
  \put(6,18.5){\makebox(0,0)[tl]
            {\begin{tabular}{ll}
               b) & A domain $\cal D$,\\
                  & the interior of\\
                  & a cut. $\psi\epsilon{\cal H}_{\cal D}$
             \end{tabular}}}
  \put(12,18.5){\makebox(0,0)[tl]
             {\begin{tabular}{ll}
               c) & The global domain \\
                  & $\cal U$. $\Psi\epsilon{\cal H}_{\cal U}$
              \end{tabular}}}
  \put(0,10.5){\makebox(0,0)[tl]
            {\begin{tabular}{ll}
               d) & Two domains \\
                  & ${\cal D}_{1}$ and ${\cal D}_{2}$,\\
                  & ${\cal D}_{1}\cap {\cal D}_{2}=\emptyset$.
             \end{tabular}}}
  \put(6,10.5){\makebox(0,0)[tl]
            {\begin{tabular}{ll}
               e) & Two domains,\\
                  & ${\cal D}_{1}\cap {\cal D}_{2}\neq \emptyset$.
             \end{tabular}}}
  \put(12,10.5){\makebox(0,0)[tl]
             {\begin{tabular}{ll}
                f) & Two domains,\\
                   & ${\cal D}_{1}\subset {\cal D}_{2}$.
              \end{tabular}}}
  \put(1.5,2.5){\makebox(0,0)[tl]
             {\begin{tabular}{ll}
                g) & Local and global\\
                   & domains $\cal D$ and $\cal U$,\\
                   & minimum needed for \\
                   & any measurement.
              \end{tabular}}}
  \put(10.5,2.5){\makebox(0,0)[tl]
              {\begin{tabular}{ll}
                 h) & Three domains\\ 
                    & ${\cal D}_{L}$, ${\cal D}_{R}$, and $\cal U$,\\
                    & possibly needed to\\
                    & describe EPR-type\\
                    & measurements.
               \end{tabular}}}
\end{picture}
  \caption{The observer/environment cut \protect}
  \label{cut}
\end{figure}
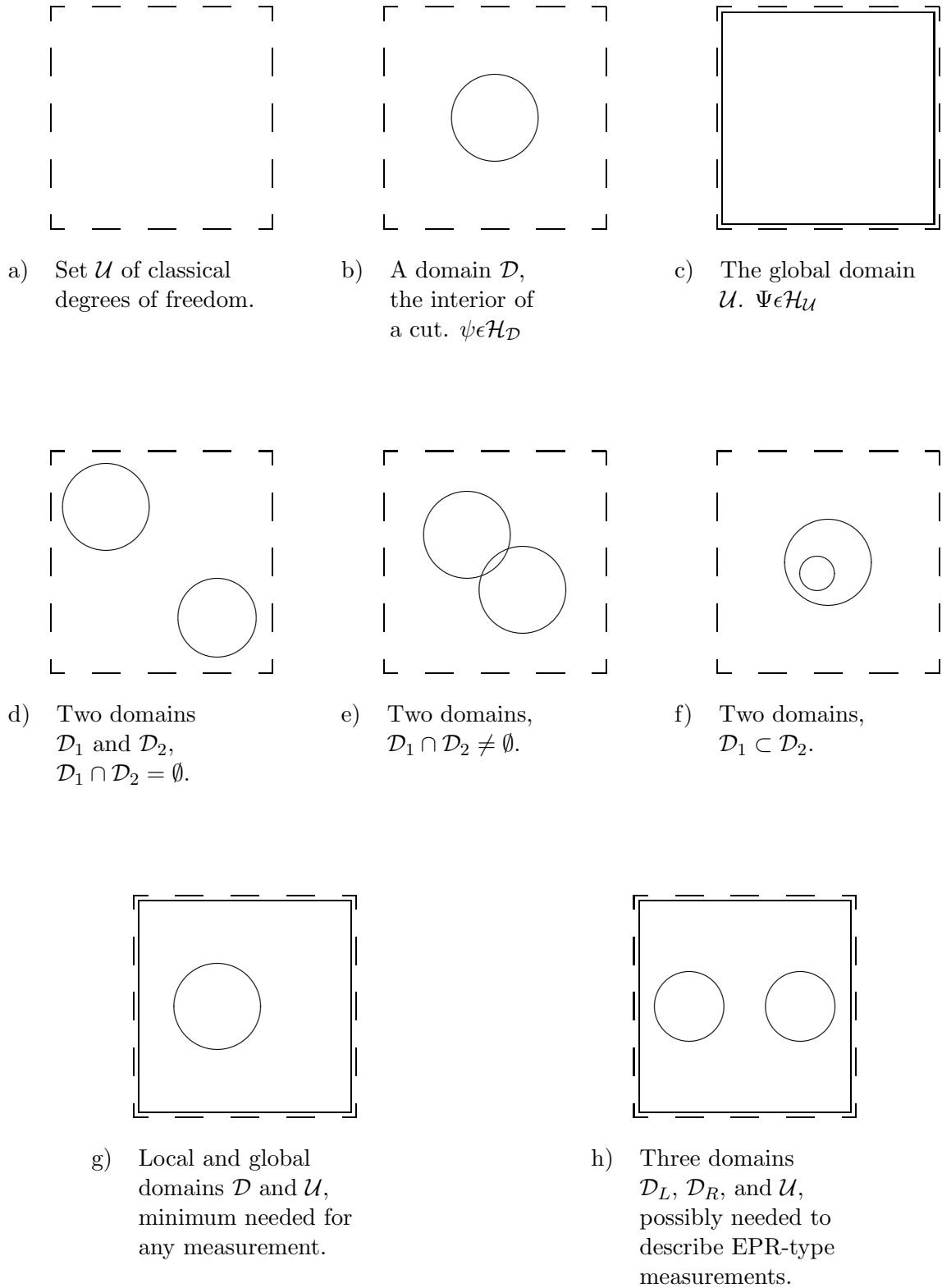

      MOM is based on both structural and dynamical changes to standard
      quantum theory: Here we present the structural and conceptual elements,
      leaving the dynamics for Part I below. 
      It is assumed that there are {\em multiple observers}, each associated 
      with a specific {\em cut}. To understand the cut, see Figure~\ref{cut}.
      First consider(a) the set $\cal U$ of all classical degrees of freedom.
      A cut(b) is an arbitrary partition of this set into an interior domain
      $\cal D$ and the exterior, the environment, which is 
      its complement ${\cal U}-{\cal D}$.
      The interior degrees of freedom form the foundation for a Hilbert space
      ${\cal H}_{\cal D}$, of which the observer, the quantum
      state $\psi$ is a member. The location of the cut is arbitrary, and it
      defines a set of classical degrees of freedom which is somehow special.
      MOM, is its present form, does not attempt to prescribe the
      choice of cuts, or explain why some are {\em active},
      i.e. have quantum states $\psi$ associated with them, and some are not.
      The usual quantum state of the universe is recovered(c) by allowing
      the interior of the cut to be equal to $\cal U$. In this work, we assume
      this cut to be active, so that the usual state in orthodox
      quantum theory is always present. In the absence of other active cuts,
      it would evolve normally, under SE. Any number of cuts may be active.
      The general relationship between
      two arbitrary cuts is depicted in (d,e and f): They may be disjoint, 
      overlap partially, or may be contained one inside the other. If they are
      disjoint, they do not affect each other directly at all, as they
      share no classical degrees of freedom. Only if they overlap partially 
      or completely will there be the potential for discrepancy between the 
      information contained in the two states, and then their evolution
      will not be independent: I.e. they will `interact'. Apparently, the
      simplest case that is non-trivial is (g), wherein there is the 
      {\em global}\/ state $\Psi$, which is always assumed to exist, and a 
      {\em local}\/ state $\psi$, associated with an active cut.
      This will be enough to describe a normal quantum measurement.
      It is possible to include more active cuts. For instance,
      results of EPR-type experiments may some day deviate from BP
      in such a way, that the simple picture, using just one local state 
      and the global state, is not able to describe the situation.
      In that case(h), we could introduce {\em two} local states, one
      associated with each detector array. The set of all possible numbers and
      arrangements of active cuts is enormous. 

      Let us now preview the results of MOM. If there is a single
      observer the evolution of its state vector is given as usual by SE:
      This is the standard quantum theory,
      minus collapse of the wavefunction. A single observer in isolation
      apparently experiences nothing but the uniform passage of time. 
      If there are two observers $A$ and $B$ we take
      ${\cal D}_{A}\subset {\cal D}_{B}={\cal U}$.
      This is the simplest case in which measurements can occur.
      The state $\psi$ in the small Hilbert space acquires an effective
      Hamiltonian, but also jumps at discrete times.
      These jumps are associated with the tendency for the full Hamiltonian
      to `decohere' the state $\psi$.
      The state $\Psi$ in the large Hilbert space just sees the
      Hamiltonian including all interactions. In addition, however, both states
      experience a new kind of interaction, which occurs between observers and
      not between degrees of freedom: This is attractive, always tending to
      make the two states compatible. Eventually both $\psi$ and
      $\Psi$ fall into compatible states, eigenstates of the
      measured operator. This `observable' is not arbitrary but is
      determined by normal interactions which cross the cut, i.e. involve
      classical degrees of freedom inside and outside of the cut domain
      ${\cal D}_{A}$.
      If there are three observers $A$, $B$, and $C$, their domains may
      overlap in a complicated way, or may be nested 
      (${\cal D}_{A}\subset {\cal D}_{B}\subset {\cal D}_{C}={\cal U}$),
      or parallel 
      (${\cal D}_{A}\subset {\cal D}_{C}={\cal U},\;
        {\cal D}_{B}\subset {\cal U},\;
        {\cal D}_{A}\cap {\cal D}_{B}=\emptyset$). 
      The latter case may
      describe an ordinary measurement with two distinct detectors, 
      such as EPR, Schroedinger's cat, or Wigner's friend.
      For simplicity we restrict ourselves to the two-observer case,
      as it already contains the elements needed to assess the merits of the
      model.

\part*{I: Dynamics}
\addcontentsline{toc}{part}{Part I: Dynamics}
\begin{quotation} \footnotesize
    I am, in fact, rather firmly convinced that the essential statistical
    character of contemporary quantum theory is solely to be ascribed to
    the fact that this (theory) operates with an incomplete description
    of physical systems\ldots
    Assuming the success of efforts to accomplish a complete description,
    the statistical quantum theory would, within the framework of future 
    physics, take an approximately analogous position to the statistical 
    mechanics within the framework of classical mechanics. I am rather
    firmly convinced that the development of theoretical physics will be
    of this type, but the path will be lengthy and difficult.
    {\em Albert Einstein} \cite[pp.666-672]{Schlipp}

    \ldots, in quantum mechanics, we are not dealing with an arbitrary
    renunciation of a more detailed analysis of atomic phenomena, but
    with a recognition that such an analysis is {\em in principle}
    excluded \ldots
    As regards the specification of the conditions for any well-defined
    application of the formalism, it is moreover essential that the
    {\em whole experimental arrangement} be taken into account \ldots
    it is decisive to recognize that, {\em however far the phenomena
    transcend the scope of classical physical explanation, the account of
    all evidence must be expressed in classical terms} \ldots
    This crucial point \ldots
    implies the {\em impossibility of any sharp separation between the
    behavior of atomic objects and the interaction with the measuring
    instruments which serve to define the conditions under which the phenomena
    appear} \ldots 
    As regards the specification of the conditions for any well-defined
    application of the formalism, it is moreover essential that the
    {\em whole experimental arrangement} be taken into account.
    {\em Niels Bohr} \cite[pp.200-241]{Schlipp}

    You believe in a dice-playing God and I in perfect laws in the world 
    of things existing as real objects, which I try to grasp in a wildly
    speculative way. {\em Albert Einstein} \cite[p.176]{Schlipp}
\end{quotation}

      The dynamics of the various states is assumed to occur in discrete time, 
      because of both theoretical and practical bias. On the theory side,
      we expect that time scales below a certain duration $\Delta t$, normally
      taken to be the Planck time, are not properly described
      by any theory of physics presently available, but that they are
      of primary importance in the measurement process.
      Therefore, since MOM is not to be considered as
      a fundamental theory of nature, it summarizes sub-$\Delta t$ dynamics
      as discrete maps of the states, occurring during succesive cycles of
      duration $\Delta t$. On the practical side, it is often
      easier to implement evolution by difference, rather than differential
      equations. Also, chaotic dynamics, which are a {\em necessary}
      ingredient of MOM in the measurement regime, are easier to produce at
      low dimensionality: There are no chaotic continuous evolution equations
      for real dimension less than three. During each cycle, there is in 
      addition to the usual action of the Hamiltonian, a decoherence-induced
      mapping of the local state vector(s), and a universal attraction between
      all pairs of observers. In this section, we describe the different
      components of the dynamics. There is the usual Hamiltonian evolution,
      the chaotic map, and a new kind of interaction between different
      observers. Each part of the dynamics in turn plays its role in the
      {\em dynamical cycle}.

\section{Effective Hamiltonian}
      It is reasonable to extract as much information as possible about the
      dynamics from the complete Hamiltonian of the system. This will help
      to maintain the distinction between ordinary physical properties of the
      system and new, model-related parameters. First we note that this global
      Hamiltonian is still applicable to $\Psi$. The effective Hamiltonian for
      $\psi$ can be obtained as follows. We construct a projection operator in
      the global Hilbert space from $\psi$ which is the product of a pure
      state operator over the local space times the identity over the
      environment space:
\begin{equation}
      {\cal P}=\psi \psi^{\dagger} \otimes \sum_{a=1}^{A} \proj{a}{a}\;. 
\label{eq:PROJ}
\end{equation}
      We then apply the standard Schroedinger evolution to this operator,
      using the full Hamiltonian, for a time interval $\Delta t$,
      and keep only those terms of order $\Delta t$. 
      The result is that $\psi$ changes but the general, factorizable form
      prescribed above does not. 
      We can then uniquely define an operator which will reproduce
      this evolution when inserted into SE for $\psi$.
      The result is the basis-independent effective Hamiltonian, or
      self-energy:
\[ \hat{H}_{eff} = \sum_{i=1}^{N} V_{i} \proj{i}{i} \; , \;\;\;
                 V_{i}= \frac{1}{A} \sum_{a=1}^{A} H_{ia} \; .  \]
      In this work this is assumed to vanish identically. It is
      constant in time if the global Hamiltonian is constant as usual.
      Note that this was derived by making the {\em no-information}
      assumption: The local state has no information whatsoever about
      the environment. If the same procedure were carried out, starting
      from the global state, $\hat{H}_{eff}$ might be different. In that case,
      it might not be possible for the global and local states to come 
      together in compatible states. 
      In the following, we assume this is not a problem.

\section{Chaotic Dynamical Map}
      Here we discuss the aspect of the dynamics which leads to chaotic
      behavior in certain cases. This stems from the decohering interactions
      assumed present in the Hamiltonian, as is represented by a discrete-time
      map. First we describe the decoherence effects, which would be present
      in a standard description of the local subsystem, and then define the
      map with a somewhat more general approach.
      The general behavior of the system under the mapping is briefly sketched.

\subsection{Decoherence}
      Let us consider the terms of order $(\Delta t)^2$, which were
      ignored above. After subtracting the self-energy contributions
      (which will be assumed to vanish from now on), we are left with terms
      which violate the assumption of factorizability. In other
      words, the interactions between the local degrees of freedom and the
      environment tend to decohere the local state $\psi$. But this is 
      contrary to the basic assumption that $\psi$ remains pure at all times.
      It is natural to postulate that some mechanism acts to
      prevent decoherence, opposing the effect of the interactions. Although
      the details of such a mechanism are beyond the scope of this model,
      we can make some observations. First, this effect vanishes if an attempt
      is made to describe it as occurring continuously in time: This is
      because it is absent to order $\Delta t$. Hence we will describe it as
      occurring at discrete time intervals. Second, the function describing
      the amount of decoherence of $\psi$ is mathematically identical,
      under suitable symmetry assumptions, to what one might guess from first
      principles(see below).
      In particular, the decohering effect vanishes near the pointer-basis
      states. Thus we are led to define a second element of the dynamics for
      $\psi$, a discrete-time map, which is chaotic if the decohering
      interactions are strong enough. These maps were previously investigated 
      numerically\cite{thesis}:
      In the chaotic regime, in the absence of competing dynamical effects,
      they lead to highly symmetrical distributions for $\psi$
      which are concentrated near the pointer-basis states. In effect, $\psi$
      undergoes a process of {\em chaotic decoherence} over time, as opposed
      to the usual decoherence which is associated with an ensemble of systems
      at a given time. 
      Note that the global state does not suffer decoherence in any way from
      the Hamiltonian, and hence is not affected by any such dynamical map.

      Specifically, let us begin at time $t=0$ with a global density
      matrix $\hat{D}$, derived from $\psi$, which is a multiple of the
      projector $\cal P$ defined above(\ref{eq:PROJ}).
      This $\hat{D}$ will evolve, under the usual SE, into $\hat{D}(t)$, 
      and will at time $\Delta t$ no longer be a multiple of a projector, but
      will have eigenvalues other than 0 and 1.
      At time $\Delta t$ we define a small decoherence parameter ${\cal Z}$,
      such that, when $\hat{D}(\Delta t)$ is reduced to a local density
      matrix, its largest eigenvalue is $1-{\cal Z}$.
      We find ${\cal Z}$ in terms of the Hamiltonian:
\begin{equation}
    {\cal Z} =\frac{1}{2} (\frac{\Delta t}{\hbar} )^{2} 
       \sum_{ij} X_{ij} p_{i}p_{j} \; , \;\; 
       X_{ij}=\frac{1}{A} \sum_{a} (H_{ia}-H_{ja})^{2}\; , \;\; 
                 p_{i}=|s_{i}|^{2} \; .
\end{equation}
      A more symmetric form is achieved by making the assumption
\[ X_{ij}\rightarrow (1-\delta_{ij} )E^{2} \; ,  \]
\begin{equation}
         {\cal Z} \rightarrow \frac{1}{2} (\frac{E\Delta t}{\hbar} )^{2}
         \sum_{i} p_{i} (1-p_{i}) \; .
\end{equation}
      We can then solve the eigenvalue problem to find the eigenstate 
      $\ket{\phi}$ corresponding to ${\cal Z}$. If, approximately,
      $\ket{\phi} \approx \ket{\psi}+\ket{\psi_{\perp}}$, then
\begin{equation}
    \ket{\psi_{\perp}} = -\frac{1}{4} \ket{\nabla_{\psi,\perp}
                 {\cal Z}}:{\rm not \;normalized}\; ,
\end{equation}
      where $\nabla_{\psi,\perp}$ 
      is the component of the gradient operator in the 
      local Hilbert space normal to $\ket{\psi}$. 
      We define the map to be a rotation of $\ket{\psi}$ in the direction of 
      $\ket{\psi_{\perp}}$,
      through an angle equal to the norm of $\ket{\psi_{\perp}}$.
      If the decohering interactions are smaller than
      $\Delta E\equiv \hbar /\Delta t$, ${\cal Z}$ remains small,
      and this procedure does not lead to the required chaotic map.
      For the map to be truly chaotic, we require 
      $t_{E}\equiv \hbar /E\ll \Delta t$: The decoherence time must be much
      less than the undetermined cyclic time scale. In what follows we will
      use $\cal Z$ as defined above, even when it is {\em not} small.

\subsection{Abstract definition of the decoherence map}
     We first define ${\cal Z}$, the generating function for the map, using
     a physically-motivated set of operators. Then we define the map in the
     general case, note the result for the case, $N=2$, 
     and continue for larger values of $N$.

     The pointer basis is assumed to have special significance during the
     measurement process, so we will use it as the basis in which to write
     the state vector $\psi$. For convenience, we will use a mixed
     braket-vector notation:
\begin{eqnarray*}
  \psi&:&\ket{\psi}=\sum_{i=1}^{N}\psi_{i}\ket{i}\;,\\
      &:&\hat{R}=\sum_{i=1}^{N}R_{i}\hat{R}_{i}\;,
\end{eqnarray*}
\[ \psi_{i}\equiv R_{i}e^{i\theta_{i}}\;,\;\;
        p_{i}=\nsq{\psi_{i}}=R_{i}^{2}\;,\;\;
                          R_{i}\in\Re\;.            \]
     The basic operators, diagonal in this basis are 
     $\hat{\theta}_{i}=\proj{i}{i}$, and their variances in the state are
\[ \Delta_{i}^{2}\equiv \matel{\psi}{\hat{\theta}_{i}^{2}}{\psi}
     -\nsq{\matel{\psi}{\hat{\theta}_{i}}{\psi}}=p_{i}(1-p_{i})\;.  \]
     A single operator is fine in the $N=2$ case, but theoretical bias
     suggests that the relevant set of operators, in the measurement of
     a larger system, should be a {\em complete set of commuting operators}.
     We will use the identity operator, with zero variance, and the set
\[ \hat{K}_{c}=\sum_{i}q_{ci}\hat{\theta}_{i}\;,\;\;c=1,2,\ldots,N-1\;. \]
     The variances and ${\cal Z}$ become
\[ \Delta_{c}^{2}=
   \sum_{i}\nsq{q_{ci}}p_{i}-\sum_{ij}q_{ci}^{\ast}q_{cj}p_{i}p_{j}\;,  \]
\[ {{\cal Z}}\equiv\sum_{c}\Delta_{c}^{2}
        =\sum_{i}K_{ii}p_{i}-\sum_{ij}K_{ij}p_{i}p_{j}\;,  \]
\[ K_{ij}\equiv \sum_{c}q_{ci}^{\ast}q_{cj}\;.  \]
     All operators in the set already commute. The condition of completeness is
     $K_{ij}=K_{i}\delta_{ij}$. For further symmetry, we set $K_{i}=K$, 
     and the function ${\cal Z}$ attains the form:
\begin{equation}
    {{\cal Z}}=K\sum_{i}p_{i}(1-p_{i})\;.   \label{eq:SYMZ}
\end{equation}
     Any map generated from this function will have a symmetric distribution,
     if no bias is introduced from another source. Also, the 
     minima(i.e. zeroes) of ${\cal Z}$ are at the pointer-basis states.

     The function ${\cal Z}$ does not depend on the phases $\theta_{i}$, so
     the vector notation is convenient. I.e., only the $R_{i}$ will be 
     affected by the map, including changes of sign. 
     We define the gradient of ${\cal Z}$ in the space of $R_{i}$,
     then the component of the gradient which is tangent to the unit sphere.
\[ \frac{\partial{{\cal Z}}}{\partial R_{i}}=2KR_{i}(1-2R_{i}^{2})\;, \]
\[ \vec{\nabla}{{\cal Z}}=
     \sum_{i}\hat{R}_{i}\frac{\partial{{\cal Z}}}{\partial R_{i}}=
     2K\sum_{i}R_{i}(1-2R_{i}^{2})\hat{R}_{i}\;,  \]
\[ \hat{R}\cdot \vec{\nabla}{{\cal Z}}=2K\sum_{i}R_{i}^{2}(1-2R_{i}^{2})\;, \]
\begin{eqnarray*}
  \vec{\nabla}_{\bot}{{\cal Z}}&\equiv&\vec{\nabla}{{\cal Z}}
                                  -\vec{\nabla}_{\|}{{\cal Z}}\;,\\
  &=&\vec{\nabla}{{\cal Z}}-\hat{R}(\hat{R}\cdot \vec{\nabla}{{\cal Z}})\;,\\
  &=&4K\sum_{i}R_{i}(S_{4}-R_{i}^{2})\hat{R}_{i}\;,
                  \;\; S_{4}=\sum_{j}R_{j}^{4}\;.
\end{eqnarray*}
     This is zero if and only if all non-zero components $R_{i}$ have the same
     magnitude. This occurs only at the pointer-basis states, and at other,
     partly symmetric points: These are the fixed points of the map.
     In the case of chaotic maps, when $K\gg 1$, the pointer-basis states are
     unstable or marginally unstable, while the others are absolutely unstable.
     Once the tangent gradient has been defined, we produce a unit-vector
     $\hat{G}$ in the same direction, and define the map as a rotation.
     The plane of rotation is spanned by $\hat{R}$ and $\hat{G}$, and the 
     angle by which the state $\hat{R}$ is to be rotated is related to the
     magnitude of the tangent gradient.
\begin{eqnarray}
  \hat{G}&\equiv&
     \|\vec{\nabla}_{\bot}{{\cal Z}}\|^{-1}\vec{\nabla}_{\bot}{{\cal Z}}\;,\\
  &=&(S_{6}-S_{4}^{2})^{-1/2}\sum_{i}R_{i}(S_{4}-R_{i}^{2})\hat{R}_{i}\;,
  \;\;S_{6}=\sum_{j}R_{j}^{6}\;. \nonumber
\end{eqnarray} 
\begin{equation}
    {\rm map}:\hat{R}\rightarrow
       \hat{R}'=\hat{R}\cos\theta+\hat{G}\sin\theta\;,\;\;
    \theta\equiv|\vec{\nabla}_{\bot}{{\cal Z}}\| \;.
\end{equation}
     The constant $K$ determines the strength of the map. The fixed points
     are just the pointer-basis states and the other places where the tangent
     gradient vanishes, as noted above.
\subsection{Mapping results}
     In case $N=2$, there is only one independent variable in the map.
     Let this be the angle $\phi\epsilon[0,4\pi]$:
\[ R_{1}^{2}\equiv \sin^{2}(\phi/2)\;,\;\; R_{2}^{2}=\cos^{2}(\phi/2)\;,\;\;
     {{\cal Z}}=(K/2)\sin^{2}\phi\;.  \]
     The map becomes
\[ \phi\rightarrow\phi'=\phi+k\cos\phi\sin\phi\equiv f(\phi)\;, \]
     where the new constant $k$ replaces $K$, and is also arbitrary 
     and real. The derivative of $f$ is
\[ df/d\phi=1+k[\cos^{2}\phi-\sin^{2}\phi] \;. \]
     We are most interested in the behavior near the pointer-basis element,
     i.e. for $\phi\ll 1$:
\[ df/d\phi\approx 1+k\;. \]

     A `hopping' method was used\cite{thesis}
     to obtain the Liapunov exponent of the map
     and the invariant distribution in the chaotic regime. The Liapunov
     exponent becomes positive for large enough values of the map strength $k$,
     and grows like the log of $k$. The invariant distribution starts as a
     small island around the maximum of the function ${\cal Z}$, and grows to
     encompass the entire space for larger values of $k$. This distribution
     becomes more uniform as $k$ is increased. Similar results were
     obtained for $N=2,3,4$.

\section{The Shift Map}
     This mapping is used to change the form and statistical properties
     of the discrete map induced on the local Hilbert space ${\cal H}$ 
     by the decohering interactions, which we will call the {\em bare} map.
     The bare map
     has some of the properties which we require for the model of
     quantum measurement processes, but it has some shortcomings. Namely, it
     leads to a distribution for the local state which is uniform over a 
     subspace of $\cal H$, determined by the initial state: 
     Any point on this subspace is visited by the local state just as often.
     This means the pointer-basis states have no privileged status, and the
     random probability walk of the global state will not yield
     the correct probabilities for outcomes of measurements, even
     approximately($\S 7$). But this distribution can be altered
     essentially at will, by use of the appropriate {\em shift map}.

\subsection{Effective map and invariant density}
     The bare map represents the effect of the decohering interactions
     on the local state, and is executed after every cycle $\Delta t$.
     The map is represented by Figure~\ref{bmap},
     where $\phi\epsilon {\cal H}$, $\phi'=f_{0}(\phi)$, 
     $f_{0}$ is the bare map, and $D_{0}$ is the distribution over
     $\cal H$ associated with $f_{0}$.
\begin{figure}[!htb]
 \centering
 \setlength{\unitlength}{1in}
\begin{picture}(6,1)(-.3,0)
  \put(1.3,.45){$\phi$}
  \put(1.6,.5){\vector(1,0){1.4}}
  \put(2.2,.6){$f_{0}$}
  \put(2.1,.3){$(D_{0})$}
  \put(3.2,.45){$\phi'$}
\end{picture}
 \caption{Bare map \protect}
 \label{bmap}
\end{figure}

     To change the map without altering its fixed points, we can use another
     function on $\cal H$, which must have an inverse: an isomorphism $g$.
     The resulting complete or effective map $f$ is represented by
     Figure~\ref{emap}, where $\psi \epsilon {\cal H}$, 
     \mbox{$\psi'=f(\psi)=g\circ f_{0}\circ g^{-1}(\psi)$},
     and $D$ is the true distribution over $\cal H$, given by
\begin{equation}
     D(\psi)=D_{0}(\phi)\| \nabla_{\psi,\bot}g(\phi)\| ^{-1}, \;\;
             \phi=g^{-1}(\psi)\;, 
\end{equation}
     where the gradient of $g$ is with respect to $\psi$, 
     tangent to ${\cal H}$ because of the constraint $\nsq{\psi}=1$,
     and the double bars around $\nabla$ indicate taking a determinant 
     of this multi-dimensional derivative. 
\begin{figure}[!htb]
 \centering
 \setlength{\unitlength}{1in}
\begin{picture}(6,2)(-.30,0)
  \put(1.3,1.65){$\phi$}
  \put(1.6,1.7){\vector(1,0){1.4}}
  \put(2.2,1.8){$f_{0}$}
  \put(2.1,1.5){$(D_{0})$}
  \put(3.2,1.65){$\phi'$}
  \put(1.4,.6){\vector(0,1){1}}
  \put(1.1,1.1){$g^{-1}$}
  \put(3.3,1.6){\vector(0,-1){1}}
  \put(3.4,1.1){$g$}
  \put(1.3,.45){$\psi$}
  \put(1.6,.5){\vector(1,0){1.4}}
  \put(2.2,.6){$f$}
  \put(2.1,.3){$(D)$}
  \put(3.2,.45){$\psi'$}
\end{picture}
 \caption{Effective map \protect}
 \label{emap}
\end{figure}

     One can see the effect of $g$ by imagining the distribution $D_{0}$
     is represented by a multitude of points $\phi$, and allowing each of them
     to be moved by $g$ to new locations. The resulting distribution is $D$.
     Now, we want the distribution to be more sharply concentrated near
     the pointer-basis states, where the decoherence generating function
     ${\cal Z}$ is minimum ($\S 4$). Also, we want to define $g$ 
     to depend only on ${\cal Z}$, without introducing any unnecessary 
     parameters. We are led to adopt the following prescription.
     The function $g$ is the result of integrating an ordinary differential
     equation(with respect to a dummy variable $s$) over a unit interval,
     with $\phi$ as initial value and $\psi$ as final value. The equation
     is
\begin{equation}
   \frac{d}{ds}\ket{X}=-\ket{\nabla_{\psi,\bot}{\cal Z}} \;,\;\;
               \ket{X(s=0)}=\ket{\phi}\;, \;\;
                  \ket{X(s=1)}=\ket{\psi}\;.
\end{equation}
     This form will ensure that $D$ is concentrated near the minima of
     ${\cal Z}$ when ${\cal Z}$ itself is large, i.e. when the decohering
     interactions are strong, and the bare map is truly chaotic.
     However, in the limit of small ${\cal Z}$, the inverse map $g^{-1}$ 
     becomes identical to $f_{0}$, so $f\rightarrow f_{0}$. Therefore we will
     mostly consider the strongly chaotic regime in what follows.

\subsection{Specific results for N=2}
     In the special case $N=2$, we can derive relatively simple specific
     formulas to help us understand the effect of the shift map $g$:
     These were useful in numerical work\cite{thesis}.
     In the cases of interest, ${\cal Z}$ has such symmetry that both
     $f_{0}$ and $g$ are one-dimensional maps. Let $x$ be the variable 
     before the shift map(like $\phi$), and $z$ the variable after the
     map(like $\psi$). We consider the behavior of the maps near a 
     fixed point(a pointer-basis state)
     which we assume to be zero: \mbox{$f_{0}(0)=g(0)=0$}.
     Near this state, $g(x)\approx ax$, where $a$ is arbitrary.
     The result is that if $f_{0}(x)\approx kx:$, then $f(z)\approx kz$.
     I.e., the map behavior near the pointer-basis states, and the capture
     process discussed in $\S 7.2$, are unchanged by the shift map.

     Now let us take the specific, standard form of the decoherence-generating
     function ${\cal Z}$, and derive the actual maps and distributions.
\[ {\cal Z} =(1/2)R^{2}p(1-p), \;\;R\equiv E\Delta t/\hbar\;\;, \]
\[ \ket{\nabla_{\psi,\bot}{\cal Z}}=R^{2}\sqrt{p(1-p)}
                              (\sqrt{1-p}\ket{1}-\sqrt{p}\ket{2})\;\;, \]
\[ \ket{\psi}\equiv \sqrt{p}\ket{1}+\sqrt{1-p}\ket{2}\;\;. \]
     Using the general form for the shift flow equation:
\[ \frac{d}{ds}\ket{\psi}=-\ket{\nabla_{\psi,\bot}{\cal Z}}\;. \]
     we get
\[ \frac{dp}{ds}=-\lambda(1-2p)p(1-p)\;,\;\lambda=2R^{2}\;. \]
     We are interested most in the behavior near the pointer-basis
     states, e.g. near $p=0$. There, this becomes 
     $dp/ds\simeq -\lambda p$, so the shift map is linear.
     For arbitrary integration limit 
     $s$(normally, $s=+1\;$ for the direct map, $s=-1\;$ for the inverse map),
     the map is $p_{0}\rightarrow p(s)=e^{-\lambda s}p$.

     We can solve the equation explicitly to obtain the shift map for any
     $p$, with the aid of some variable changes:
\[ p\equiv \sin^{2}\theta, \;\;
      z\equiv \tanh^{-1}(\cos \varphi), \;\;\varphi \equiv 4\theta, \]
\[ \frac{dz}{ds}=\lambda/2\;, \;\;z(s)=z_{0}+R^{2}s\;, \]
\[ p(s)=\sin^{2}(\frac{1}{4}\cos^{-1}\tanh(z_{0}+R^{2}s))\;. \]
     We can also solve for the bare map explicitly. In terms of $\theta$,
     it is defined as a rotation in the direction of the gradient
     $\ket{\nabla_{\psi,\bot}{\cal Z}}$, through an angle of magnitude 
     $\| \;\ket{\nabla_{\bot}{\cal Z}}\;\|$. 
\begin{eqnarray*}
  \Delta \theta&=&R^{2}\| \cos 2\theta \|
                                  (\frac{1}{4}\sin^{2}2\theta)^{1/2}\;,  \\
          &\simeq&R^{2}\theta\;,\;\; \theta \simeq 0\;.
\end{eqnarray*}
     I.e., we get linear behavior:
\begin{eqnarray*}
  \Delta \theta&\simeq&R^{2}\theta\;,  \\
       \Delta p&\simeq&(1+R^{2})^{2}p\simeq R^{4}p\;,\;\; 
                          {\rm in \;the \;chaotic \;regime.} 
\end{eqnarray*}
     The variable $\varphi$ is the most transparent:
\begin{equation}
    \varphi \rightarrow \varphi'=\varphi+R^{2}\sin \varphi \;.
\end{equation}
     The distribution is nearly constant over $\varphi$, so for $p$ is
     approximately
\[ D(p)=\frac{2}{\pi}/\sqrt{p(1-p)}\;. \]
     In terms of the transformed variable $p'$, the shift map changes 
     this to
\[ D'(p')=K\frac{2}{\pi}/\sqrt{p'(1-p')}\;, \]
\[ K\equiv\frac{\sqrt{1-\rho^{2}}}{1-\rho(1-8p'(1-p'))}\;,\;\;
   \rho\equiv\tanh R^{2}s\;.                                   \] 
     We concentrate on the factor $K$. Since we are in the chaotic regime
     by assumption, $R^{2}s\gg 1$, so $\rho \simeq 1-2e^{-2R^{2}s}$.
     The factor becomes
\[ K\simeq e^{R^{2}s}/(1+e^{2R^{2}s}4p'(1-p'))\;. \]
     This is about one when $p'\simeq \bar{p}=(1/4)e^{-R^{2}s}$, a very
     small value. Defining $p'\equiv a\bar{p}$, the factor $K$ behaves thus:
\begin{eqnarray*}
     a\ll e^{-R^{2}s}&:&K\simeq e^{R^{2}s} \gg 1\;,  \\
  a\simeq e^{-R^{2}s}&:&K\simeq (1/2)e^{R^{2}s} \gg 1\;, \\
                  a=1&:&K=1\;, \\
               a\gg 1&:&K\simeq 1/a\ll 1\;. 
\end{eqnarray*}
     The net result is that, in the chaotic regime, the shift map
     causes the distribution over $p$ to become severely concentrated near
     the pointer-basis states, as required by $\S 7$. The spread about each of
     these states should be no more than $\bar{p}$, which decreases 
     {\em exponentially} with increasing decohering interaction strength.

     Finally we summarize the explicit forms of the maps for the $N=2$ case,
     with ${\cal Z}$ having the standard form and $s>0$:
\begin{equation}
     f=g\circ f_{0} \circ g^{-1}:\varphi \rightarrow \varphi'\;,
\end{equation}
\begin{eqnarray*}
  g^{-1}:\varphi&\rightarrow&\varphi_{1}=
                    \cos^{-1}\tanh(z-R^{2}s)\;,  \\
   f_{0}:\varphi_{1}&\rightarrow&\varphi_{2}=
                    \varphi_{1}+R^{2}\sin \varphi_{1}\;,  \\
       g:\varphi_{2}&\rightarrow&\varphi'=
                    \cos^{-1}\tanh(z_{2}+R^{2}s)\;,
\end{eqnarray*}
     where $z$ and $z{_2}$ correspond to $\varphi$ and $\varphi_{2}$,
     respectively.

\section{Observer Interactions}
      This element of the dynamics is completely new, consisting of an
      attraction between pairs of states which, in the absence of other
      effects,
      results in the following. The global state $\Psi$ tends toward a form
      which is factored between local and environment degrees of freedom.
      That factor containing the local information is the same as the local 
      state $\psi$. Hence there is effectively only one description of the
      system, $\Psi$. However, this is true only asymptotically for large
      times, so that at any finite time there remains some information in
      $\psi$ which differs slightly from $\Psi$. Symbolically, we have
\begin{equation}
  \ket{\psi} \rightarrow \ket{\psi'}\;,\;\;
  \ket{\Psi} \rightarrow \ket{\Psi'}=\ket{\psi'}\otimes
                     \sum_{a=1}^{A}C_{a}\ket{a} \; .
\end{equation}

\subsection{Summary of relevant features}
      For simplicity we describe this effect by a
      discrete-time map which is geometrically motivated, so that only
      the strength of the attraction must be defined, given the two states 
      involved. The result on $\Psi$ is
\begin{equation}
      \ket{\Psi} \rightarrow \ket{\Psi'}=A_{1}{\cal P}\ket{\Psi}
                   +A_{2}(1-{\cal P})\ket{\Psi} \; ,
\end{equation}
\[ A_{1}=1+\alpha (1-z) \; , \; A_{2}=1-\alpha z(2+\alpha (1-z))
                 \; , \; z=\matel{\psi}{\hat{Q}}{\psi} \; ,  \]
      where $\cal P$ and $\hat{Q}$ are defined 
      above(\ref{eq:PROJ},\ref{eq:GDM}).
      Here $\alpha$ represents the strength of the
      map, and is a positive real function of $z$, the overlap between $\psi$
      and $\Psi$. For the effect on $\psi$ we have
\begin{equation}
       \ket{\psi} \rightarrow \ket{\psi'}
       =a\hat{Q}\ket{\psi}+b(1-\frac{z}{y} \hat{Q})\ket{\psi} \; ,
\end{equation}
\[ y=\matel{\psi}{\hat{Q}^{2}}{\psi} \; , \;\;
                 a^{2}=(1-b^{2}(1-\frac{z^{2}}{y}))/y \; .  \]
      Here the strength parameter is taken as $b$, between 0 and 1.
      The discrete time scale $\Delta t$ is the same as above.
      We choose to make the strength of the attraction map on one
      state due to the other state depend only on the dimensionality of the
      two Hilbert spaces, and on the overlap $z$.
      Below we will see this ensures that if additional,
      irrelevant degrees of freedom are added to the system, there will be
      no significant effect on the dynamics. There, we also show, that the 
      relative strength of the two maps should equal the ratio of the 
      dimensionalities of the Hilbert spaces. This interaction has some
      formal similarities with gravitation between massive bodies: It acts
      universally between pairs of objects (state vectors), it is attractive,
      and its action on a given object depends on a measure of the object's
      `inertia', here measured by the dimensionality.
      This helps one to imagine the behavior of the state vectors under the
      influence of these, otherwise novel, interactions.

\subsection{Abstract definition of the attraction map}
     We consider the effect of one state on another state.
     First we take both states to be in the same Hilbert space,
     and then generalize by using a density matrix to represent the agent
     state. Finally we consider the stability of the map:
     If degrees of freedom are added to one or both states, 
     how is the map affected? In the process, we will discover the way
     the {\em effective} strength of the map depends on the dimensionality
     of the Hilbert spaces.

\subsubsection*{Action of state on state}
     In this case, both states belong to the same Hilbert space. It is
     most convenient to use the projector $\hat{P}=\proj{B}{B}$ 
     to represent the state B, which acts on state A. The map is defined as
\begin{equation}
   \ket{A}\rightarrow \ket{A'}=a\hat{P}\ket{A}+b(1-\hat{P})\ket{A}\;,
\end{equation}
\[ a^{2}=[1-b^{2}(1-z)]/z\;, \]
     where $z$ is the {\em overlap} of A and B,
\[ z=\matel{A}{\hat{P}}{A}=\nsq{\braket{A}{B}}\;, \]
     which is zero if they are orthogonal, and one if they coincide.
     For $b=1$, the map is weakest ($\ket{A'}=\ket{A}$), while for $b=0$, it 
     is strongest ($\ket{A'}=\ket{B}$). The fixed-point condition of $\S 7.1$,
     where the probability walk of the global state is constrained to maintain
     BP, suggests we define the constant $\beta$:
\[ 1-b^{2}\equiv z\beta\;. \]
     The map now takes the form
\begin{equation}
         \ket{A'}=\sqrt{1+\beta(1-z)}\hat{P}\ket{A}
              +\sqrt{1-\beta z}(1-\hat{P})\ket{A}\;.
\end{equation}
     The value $\beta=0$ gives the weakest map, the identity. The maximum
     value of $\beta^{2}$ is found by constraining the square roots to be
     real:
\begin{eqnarray*}
  (1+\beta(1-z)\geq 0)&\Rightarrow&(\beta\geq -1/(1-z) \geq -1)\;,\\
    (1-\beta z \geq 0)&\Rightarrow&(\beta\leq 1/z \leq +1)\;.
\end{eqnarray*}
     This gives $-1 \leq \beta \leq +1$.
     The variable of interest is $z$. It evolves according to
\begin{equation}
      z\rightarrow z'=z+\beta z(1-z)\equiv f_{\beta}(z)\;,
\end{equation}
\begin{eqnarray*}
   df_{\beta}/dz&=&1+\beta(1-2z)\;,\\
   &\stackrel{z\rightarrow 0}{\longrightarrow}&1+\beta\;,\\
   &\stackrel{z\rightarrow 1}{\longrightarrow}&1-\beta\;,
\end{eqnarray*}
     which shows that, for $\beta \rightarrow -\beta$, the effect is reversed:
     I.e. the {\em attraction} becomes {\em repulsion}. For $\beta=0$, the map
     is trivial: $z'=z$. For $\beta=\pm 1$, the map is maximally attracting
     or repulsing, especially near $z=1$ and $0$, respectively.

\subsubsection*{Action of density matrix on state}
     In general the two states involved are not in the same Hilbert space.
     Consider the global and local states of the standard measurement scenario.
     The local state can be promoted to a density matrix in the global 
     Hilbert space, or a projection operator with trace bigger than one.
     The global state can be reduced to a density matrix in the local Hilbert
     space, in general not a pure-state projector. 
     In either case, the appropriate
     generalization of the previous section involves defining the effect of
     a density matrix $\hat{Q}$ on a pure state $\psi$. Because $\hat{Q}$ is 
     not a pure-state projector, $\hat{Q}^{2}\neq \hat{Q}$, so the algebra is 
     slightly different. A convenient definition of the map is
\begin{equation}
      \ket{\psi}\rightarrow \ket{\psi'}=a\hat{Q}\ket{\psi}
                     +b(1-\frac{z}{\Gamma}\hat{Q})\ket{\psi}\;,
\end{equation}
\[ z=\matel{\psi}{\hat{Q}}{\psi}\;, \;\;
     \Gamma=\matel{\psi}{\hat{Q}^{2}}{\psi}\;,\;\;
     a^{2}=[1-b^{2}(1-z^{2}/\Gamma)]/\Gamma\;. \]
     In the same way as above, we define the strength parameter $\beta$.
     The map becomes
\begin{equation}
     \ket{\psi'}=
     \frac{z}{\Gamma}\sqrt{1+\beta\frac{z}{\Gamma}(1-z)}\hat{Q}\ket{\psi}
     +\sqrt{1-\beta z}(1-\frac{z}{\Gamma}\hat{Q})\ket{\psi}\;.
\end{equation}
     It is of interest to note the behavior of the local state under this map,
     when the density matrix is close to a pure-state projector. Let 
     $\hat{Q}$ and $\ket{\psi}$ be written in terms of a basis which includes
     this state as the member with label $i=\mu$, then define 
\[ Q\equiv \sum_{i\neq\mu} Q_{ii}\ll 1\;,\;\; Q_{\mu\mu}=1-Q\;, \]
\[ p\equiv \sum_{i\neq\mu}\nsq{\psi_{i}}\;,\;\; \nsq{\psi_{\mu}}=1-p\;, \]
\[ Q\circ p\equiv \sum_{i\neq\mu} Q_{ii}\nsq{\psi_{i}}\ll 1\;. \]
     First consider the case $Q=0$, when the density matrix is a pure-state
     projector. For small $p$, the local state almost coincides with
     that state. The map becomes, for arbitrary $p$,
\begin{eqnarray*}
  p\rightarrow p'&=&(1-\beta)p+\beta p^{2}\;,\\
  &\stackrel{\beta\rightarrow 0}{\longrightarrow}&p\;,\\
  &\stackrel{\beta\rightarrow +1}{\longrightarrow}&p^{2}\;,\\
  &\stackrel{\beta\rightarrow -1}{\longrightarrow}&2p-p^{2}\approx 2p\;.
\end{eqnarray*}
     We will consider only $\beta \geq 0$, except in $\S 9.3$.
     When $\beta=+1$, this will overcome any linear map, such as the 
     decoherence map above, even in the chaotic regime. We use that
     fact in $\S 7.2$, when considering the capture process, whereby the local
     state is held near a pointer-basis state by the GDM.
     If both $Q$ and $p$ are small but non-zero, the map is, to leading order,
\begin{eqnarray}
  p'&\approx&
    (1-\beta)p+[\beta Qp+2\sqrt{1-\beta}(1-\sqrt{1-\beta})Q\circ p]\;,\\
  &\stackrel{\beta\rightarrow 0}{\longrightarrow}&p\;, \nonumber \\
  &\stackrel{\beta\rightarrow +1}{\longrightarrow}&Qp\;. \nonumber
\end{eqnarray}
     This form will help us put a bound on $Q$, for capture to occur.

\subsubsection*{Strength parameter}
     The question, what value should $\beta$ take, is related to whether
     a projector or a density matrix should be used to define the map.
     Above we used a projector to represent the local state, boosted
     up to the bigger Hilbert space of the global state. This had a large
     trace,
     but can be rescaled down to a density matrix with trace one.  Consider
     a map defined by the matrix $\hat{M}$, and change the map by multiplying
     $\hat{M}$ by a constant, and possibly also changing $\beta$:
\begin{eqnarray*}
  \hat{M}&\rightarrow&\hat{M}'=k\hat{M}\;,\\
    \beta&\rightarrow&\beta'\;,\\
        z&\rightarrow&z'=kz\;,\\
   \Gamma&\rightarrow&\Gamma'=k^{2}\Gamma\;,\\
        a&\rightarrow&a'=k^{-1}\sqrt{[1-(b')^{2}(1-z^{2}/\Gamma)]/\Gamma}\;,\\
        b&\rightarrow&b'=\sqrt{1-k\beta'z}\;,\\
   \ket{\psi'}&\rightarrow&\ket{\psi'}'=ka'\hat{M}\ket{\psi}
           +b'(1-\frac{z}{\Gamma}\hat{M})\ket{\psi}\;.
\end{eqnarray*}
     We can regain the original map, if we let $k\beta'=\beta$. Since the
     trace of $\hat{M}$ has been multiplied by $k$, this suggests
\begin{equation}
   \beta{\rm Trace}(\hat{U})\equiv \beta_{0}:{\rm constant}\;.
\end{equation}
     Here $\hat{U}$ is whatever operator we want to use to define the map, 
     and the constant $\beta_{0}$ will be taken to be equal to one,
     which will make the maps as strong as possible. 

     Now we can see what this means, in terms of the dimensionality of the 
     Hilbert spaces. If both states are from the same space, then the map
     is maximally strong, with $\beta=1$. If the agent state is from a smaller
     space, then assuming a projector is used instead of the boosted density
     matrix, $\beta=1/R\ll 1$, where $R$ is the ratio of the numbers of 
     linearly independent states in the two Hilbert spaces 
     (i.e. their dimensionality). In practice, the map
     is defined more naturally in terms of the projector, 
     so this means the value of $\beta$ appearing in the equations above
     will be small, and the local state
     has only a weak effect on the global state. If the agent state is from
     a larger space, then also $\beta=1$, because the reduced density matrix
     used above to define the map already has trace equal to one,
     so the global state affects the local one strongly.      

\subsubsection*{General case and stability}
     Now we want to define the map in general, and see what happens,
     when degrees of freedom are added somewhere.
     We distinguish between the mapped state $\psi_{B}$ on domain $B$, 
     to which the map is being applied, and the agent state $\psi_{A}$ on
     domain $A$, which defines the map. The sets $A$
     and $B$ will be arbitrary. We define their intersection $C=A\cap B$,
     the non-overlapping parts $\tilde{A}=A-C$ and $\tilde{B}=B-C$, and their
     common environment $\tilde{U}=U-(A\cup B)$, where $U$ is the universe.
     Note that the simplified cases
     considered so far were first, $A=B$ or {\em state on state}, and then
     $A\subset B$ or {\em density matrix on state}.
     The effect of $\psi_{A}$ on $\psi_{B}$ is obtained as follows. 
     We first form $\hat{D}_{C}(\psi_{A})$, the reduced density matrix in the
     space of $C$ obtained from $\psi_{A}$. We then extend this to a density
     matrix $\hat{D}_{B}(\psi_{A})$ on the space of $B$, by first forming a
     projector and then normalizing it to unit trace. In this normalization,
     the factor $\beta$ is introduced, which is the ratio of the
     dimension of the space of $C$ to that of the space of $B$. This is
     always less than or equal to one: It controls the strength of the
     attraction map. The map is now defined as in the previous section, 
     as a density matrix acting on state.
     Now consider adding degrees of freedom to the various domains. 
\begin{description}
  \item[Add to $\tilde{A}$:] In this case, these are
    traced over while forming the reduced density matrix $\hat{D}_{C}$. If the
    `new' state $\psi_{A}$ still contains no correlation between the new
    degrees of freedom and $C$, then $\hat{D}_{C}$ does not
    change. If some correlation is included between the new degrees of
    freedom and $C$, this has an effect on $\hat{D}_{C}$, typically making
    it less coherent. Since new information has been added, this is not
    incorrect. In any event, the factor $\beta$ is unchanged.
  \item[Add to $\tilde{B}$:] In this case, the factor
    $\beta$ is affected directly: It decreases, making the map weaker.
    The `inertia' of state $\psi_{B}$ has become greater.
  \item[Add to $C$:] In this case, we have added the same
    degrees of freedom to both $A$ and $B$, so they are more nearly the same
    size. The factor $\beta$ increases, reflecting this, unless it is
    already equal to one.
  \item[Add to $\tilde{U}$:] In this case the map is completely unaffected by
    construction.
\end{description}    
 
     In conclusion, we have defined the universal attraction map, which
     represents the influence of one state upon another, and serves
     to bring their information content into agreement. There is strong
     indication, that the agent state should be represented by a trace-one
     operator, acting with universal strength $\beta_{0}$, which
     we let equal one. In particular, this means that if one of the 
     states is from a smaller Hilbert space, it will have a weaker influence 
     on the other. The mapped state should always be represented
     by a pure state in its natural Hilbert space, not including spurious
     degrees of freedom, about which there is no information. Finally, we note
     that this is a {\em consensus-building} map, which acts to decrease
     disagreement between the two states about any degrees of freedom.

\section{Collapse}
      Here we consider the basic collapse process as described by the
      dynamics of MOM. More details, enabling us to describe an actual
      measurement from beginning to end, will be supplied in $\S 8$.
      We have described three dynamical elements: the Hamiltonian, the
      decoherence map, and the attraction map. They do not act in isolation,
      but together, to define the actual dynamics of $\psi$ and $\Psi$.
      The details of how they are combined should depend on an underlying
      theory which we do not yet know. We assume that they act
      sequentially within each cyclic time interval $\Delta t$:
      First the Hamiltonian will act (including all decoherence effects)
      and then the universal attraction.

      To understand the role of collapse within the overall process,
      a brief sketch of a measurement is necessary.
      Before a measurement begins, the two states are almost in agreement
      at one of the pointer-basis states: The global state is factorized.
      In the preparation phase of every measurement there are, in addition
      to the pointer-basis interactions, other interactions which are not
      of that type, which we will call {\em active}. In the case of a
      controlled experiment, these interactions serve
      to manipulate or prepare the system into an appropriate superposition
      or mixture of pointer-basis states whose properties will be measured.
      They move the global state $\Psi$ away from its original pointer-basis
      state, introducing correlations between the local and environment
      degrees of freedom: The local state $\psi$ is basically unaffected.
      When the active interactions are effectively over there are left only
      the pointer-basis interactions. What follows is a period of time
      wherein $\psi$ undergoes chaotic evolution(if the decohering
      interactions are strong enough) and $\Psi$ acts as if under the
      influence of a random walk, while maintaining on average the
      probabilities associated with each pointer-basis state, which were 
      generated by the active interactions. Finally $\Psi$ comes close enough
      to a pointer-basis state to capture $\psi$ via the attraction map.
      Thereafter the two states coalesce asymptotically.
      Thus the system as a whole ($\psi$ and $\Psi$) starts
      near a pointer-basis state and ends near one. The probability for it to
      end near a  given state is determined solely by the form
      of the active interactions which initiated the measurement. 
      We can summarize the process as follows.
      In the initial phase, the local and global(factored)
      states, which begin very close together, are drawn apart. In the
      middle phase they undergo chaotic, (pseudo-)random motion.
      In the final phase capture of the local state by the
      global(factored) state occurs, very near a pointer-basis state which
      becomes the result of the measurement.
      We now consider random walk and capture phases in more detail.

\subsection{Probability walk}
      In the middle phase of measurement the two states are typically not close
      together. The behavior of $\psi$ is dominated by the chaotic
      map, although the attraction map does introduce a slight bias which
      breaks the symmetry of its distribution in the Hilbert space. It is
      important to consider the strengths of the chaotic map and of the
      attraction map. The first is determined by the ratio of the time
      scale $\Delta t$ to the decohering interaction time scale.
      As long as these interactions are fast enough the `chaotic' map will
      {\em actually be} in the strongly chaotic regime, characterized by
      a large Liapunov exponent and small correlation time. The second,
      the strength of the attraction map, is determined by an arbitrary
      parameter and by the degree of local coherence of the global state.
      This attraction will dominate only when the global state is almost
      totally locally coherent, near a pointer-basis state.
      Meanwhile the global state is affected by the global Hamiltonian,
      which tends to make its GDM
      diagonal in the pointer-basis, and by the attraction map, which should
      be weak as seen above. The resulting motion is a (pseudo-)random walk,
      because the local state is rapidly changing and filling in approximately
      the invariant distribution of the chaotic map. Here an important
      constraint must be observed\footnote{I.e., the `martingale' property;  
      e.g., see Feller \cite{Feller}.}: During this motion, the
      {\em diagonal elements} of the GDM, $\hat{Q}$ (\ref{eq:GDM}),
      should remain approximately constant in time, {\em on average}.

      To impose the constraint we will replace $\psi$ by the distribution over
      the Hilbert space which the chaotic map generates. This is approximated
      by a symmetric sum of $\delta$-functions at the pointer-basis states.
      Then we require that, for any $\hat{Q}$, the change in its diagonal
      components under the attraction map is zero, when averaged over the
      distribution. This will be true if
\begin{equation}
    \alpha (2+\alpha (1-z))=constant\equiv \beta ,\; 0<\beta \leq 1 \; .
\end{equation}

      The map becomes
\begin{equation}
     \ket{\Psi} \rightarrow \ket{\Psi'}=
             \sqrt{1+\beta(1-z)}{\cal P} \ket{\Psi} +
             \sqrt{1-\beta z} (1-{\cal P}) \ket{\Psi} \; .
\end{equation}
      It is instructive to see the effect on $z$ of this process alone.
      Recall that $z$ is the overlap between $\psi$ and $\Psi$, ranging
      between 0 and 1. Under repeated application of the map, 
\[ z\rightarrow z'=z+\beta z(1-z)\rightarrow \cdots  \rightarrow 1 \; . \]
      This map has fixed points at 0(unstable) and 1(stable), is the identity
      when $\beta =0$, and forces $z\rightarrow 1$ most rapidly when
      $\beta =1$. 
      For now we will assume both the spread and bias of the invariant
      distribution is negligible. Failure for this to be strictly true will
      result in deviations from BP. Also, note that in the
      above prescription, the projector $\cal P$ (\ref{eq:PROJ})
      is used, rather than the density matrix $\hat{G}$, which is the more
      appropriate object because Trace$(\hat{G})=1$.

\subsection{Capture}
      The final phase is the capture process, which gives meaning to the
      probabilities represented by the diagonal components of $\hat{Q}$.
      Normally, all fixed points of the combined attraction/chaotic
      map for $\psi$ are unstable. But if $\hat{Q}$ is very nearly coherent
      and close to a pointer-basis state, this becomes a stable fixed point,
      an attractor. Soon $\psi$ visits the neighborhood of this point and stays
      there, also holding in place $\Psi$, unless the active interactions 
      pull it away too rapidly. The two states coalesce
      asymptotically for large time. The requirement that this picture should
      be valid, however, places some constraints on the local state maps.
      First, the attraction map is linear near the pointer-basis states
      (except see below), so the chaotic map must also be linear there.
      If the power is less than one, no capture will ever occur. 
      If it is greater than one, capture will occur too soon and the 
      probabilities will be violated. Therefore, as is implicit in the above,
      only the strength of the chaotic map is arbitrary but not its power-law
      behavior. Second, the attraction map strength on the local state must
      be at or near maximum in the following sense. The strength is defined
      by the constant $\gamma$, with $b^{2}=1-\gamma z$, which ranges between
      0(weakest) and 1(strongest). For capture to occur $\gamma$ must be at
      or near 1. This suggests that if the environment degrees of freedom
      are very numerous $\gamma$ should approach one from below, yielding the
      greatest efficiency for measurement in `macroscopic' experimental
      set-ups, as we would expect.
     
      To see this, we consider the capture process in some detail. We assume
      for simplicity a two-level local system ($N\!=\!2$),
      and consider the effects
      of the chaotic and attraction maps on $\psi$, when both it and $\Psi$
      are near a pointer-basis state. We define $p_{1}=1-x$, $Q_{11}=1-P$,
      with $x$ and $P$, the probability-distances of the local and global
      states from the pointer-basis state, very small.
      The effect of the attraction map on $x$ is
\begin{eqnarray*}
  (\gamma <1) & x\rightarrow x' & \approx (1-\gamma )x \; ,  \\
  (\gamma =1) &                 & \approx f(P)x \; , 
\end{eqnarray*}
\[ f(P)=(\sqrt{1-\gamma (1-P)} \frac{1-2P}{1-P}+\frac{P}{1-P} )^{2} \; . \]
      The effect of the chaotic map on $x$ is
\[ x'\rightarrow x''\approx (1+\lambda^{2} )^{2}x' \; , \;\; 
        \lambda =\frac{\Delta t}{t_{E}} \gg 1 \; \mbox{ for chaos.} \]
      We now let both maps act in sequence, and require that $\psi$ be drawn
      even closer to the pointer-basis state. I.e., $x''<x$ when $P<P^{\ast}$
      and $x<x^{\ast}$, for some small $P^{\ast}$ and $x^{\ast}$. 
      If $\gamma =1$, capture always occurs because the attraction map
      becomes quadratic as $P\rightarrow 0$ and overpowers the linear
      chaotic map. In this case $P^{\ast} \approx \lambda^{-4}$:
      For chaotic motion, this is a very small number, as it should be if 
      large probability deviations, which are at least of order $P^{\ast}$,
      are to be avoided. 
      If $\gamma <1$, the attraction map is always linear and must compete 
      with the chaotic map: Capture will not take place if $E$, 
      the scale of decohering interactions, is large enough.
      For now we assume that $\gamma=1$, so that capture takes place.

\subsection{Measurement time}
      It is important to consider the average measurement time interval: how
      long a typical measurement requires. We can make an estimate based on
      the properties of random walks, in units of $\Delta t$.
      We first consider $z$ to be undergoing an unbiased random walk:
\begin{equation}
    z\rightarrow z\pm \beta z(1-z) \; , \; \pm ={\rm random.}
\end{equation}
      Then we transform to a variable $s$ with $\Delta s=\pm 1$, and use the
      standard result in Feller \cite{Feller}. The average measurement time is
\begin{equation}
  T_{meas}\approx (\ln[1+(\Delta t/t_{E})^{2}]^{2})^{2} \Delta t/\beta^{2}
   \approx 16(\ln[\Delta t/t_{E}])^{2}\Delta t/\beta^{2}\; .
\end{equation}
      We can see what the minimum for this average measurement time will
      be by noting that the ratio $(\Delta t/t_{E})^{2}$ is the strength of
      the map induced on the local state by the decohering interactions.
      This must be large enough for the map to be in the chaotic regime:
      $(\Delta t/t_{E})^{2}\ge \rho$, where $\rho \approx 8$.
      Hence the minimum average time for {\em any} measurement to occur is
\[ T_{meas,min}=4(\ln[1+\rho])^{2}\Delta t/\beta^{2}\;.  \]
      We can use this right away to constrain $\beta$, given some experimental
      input. If we observe a quantum measurement to occur during a time 
      interval less than $T^{\star}$, then $\beta$ must be greater than 
      $\beta^{\star}$, where
\[ \beta^{\star}=2\ln[1+\rho]\sqrt{\Delta t/T^{\star}}
                 \approx 5\sqrt{\Delta t/T^{\star}}\;.  \]
      Using $\Delta t\approx 10^{-43}s$, we consider a few possibilities.
\begin{eqnarray*}
  T^{\star}&=&10^{-23}({\rm strong \;interactions})
              \Rightarrow \beta^{\star} \approx 10^{-9}\;,  \\      
           &=&10^{-13}({\rm electromagnetism})
              \Rightarrow \beta^{\star} \approx 10^{-14}\;, \\
           &=&10^{-1}({\rm human \;perception})
              \Rightarrow \beta^{\star} \approx 10^{-20}\;.
\end{eqnarray*}
      At present, there is no definitive evidence of any quantum measurements
      occurring faster than the human perception scale, because otherwise
      we would have {\em proof}\, of the need for the `collapse of the
      wavefunction', ruling out \mbox{`no-collapse'} models. Hence the only
      bound provided is $\beta >10^{-20}$: Even this is not very firm.

      The uncertainty relation(UR) for time-energy holds that the measurement
      time should be at least as long as the decoherence time:
      $T_{meas}E\geq\hbar$. This is true automatically:
\begin{description}
  \item[$t_{E}<\Delta t$:] In the chaotic regime, the measurement time 
      is always much longer.
  \item[$t_{E}>\Delta t$:] In the non-chaotic regime, the measurement is not
      produced by the model, but is instead deferred until a later time, when
      other degrees of freedom and/or interactions become involved.
\end{description}
      Indeed, the only problem is that, in the chaotic regime where 
      measurements actually occur, $T_{meas}$ is too long. This cannot
      be helped because the model does not attempt to describe reality at 
      time scales below $\Delta t$: That is the main reason why effective 
      discrete-time maps are used. If an interaction crosses a cut {\em and}\,
      it exceeds $\Delta E\equiv\hbar/\Delta t$ sufficiently, it will cause a
      measurement, which will take longer than one might have expected. 
      From an experimental point of view, however, this is not bad:
      It would be much more difficult to observe measurement times which
      were shorter than UR allows, than times which are longer.

      To summarize, MOM results in measurements which have a minimum
      average duration. This could be used to constrain $\beta$, if suitable
      experiments verifying the existence of the collapse process were
      performed: Otherwise, we can still say $\beta>10^{-20}$, based on the
      human perception time scale.
      In MOM, measurements occur only at decoherence energies above the
      energy scale $\Delta E$, and then the measurement time exceeds the
      value prescribed by UR.

\subsection{Probability deviations}
      The frequencies for outcomes of measurements in MOM deviate from BP
      due to at least two effects.
      First there is the finite size of the region for capture of the
      local state by the global state, and second there is the fact that the
      time-averaged distribution of the local state, as it undergoes chaotic
      motion, is not perfectly and symmetrically concentrated at the 
      pointer-basis states. Here we will discuss these two effects in turn.

      The effect of the finite size of the capture region can be seen by
      returning to the basic probability walk introduced in $\S 1.3$.
      In the one-dimensional case, let the diagonal element of the
      GDM be $z$, between 0 and 1.  
      The usual BP is just $p\equiv z$, and one can check that the 
      random walk keeps this constant, on average, at each step: This was
      the constraint used in $\S 7.1$ to ensure BP would be reproduced
      by the model(approximately). The {\em true} probability $q(z)$ obeys the
      boundary conditions
\[ q(\delta)=0, \;\;q(1-\delta)=1, \;\; \delta=(\Delta t/t_{E})^{-4}\ll 1, \]
      and the self-consistency equation
\begin{equation}
         q(z)=\frac{1}{2}[q(z_{+})+q(z_{-})]\;, \;\; 
                   z_{\pm}=z\pm \Delta z\;, \;\;
                     \Delta z=\beta z(1-z).
\end{equation}
      This just says the value of $q$ at the mid-point between $z_{+}$ and
      $z_{-}$ is the average of the values at the two points. 
      The solution should be a linear function of $z$, namely
\[ q(z)=\frac{z-\delta}{1-2\delta}\;,  \]
      in which case the maximum deviation from BP, occurs at 
      $z=\delta$ and $z=1-\delta$, and has the value $\delta$. 
      The only problem with the above derivation is, that due to the finite
      size of $\beta$, the linearity cannot actually hold over a {\em finite}
      interval, when $z$ is within about $\beta \delta$ of a capture region.
      The actual solution for $q$ seems to be a series of steps, which
      tend toward a straight line, in the limit as $\beta \rightarrow 0$.
      Since $\beta$ is small by assumption, we assume that the deviation from
      linearity for $q(z)$ is no worse than $\beta \delta$, and therefore
      our estimate is approximately correct. I.e. the deviation from BP
      due to the finite size of the capture region is no worse than the size
      of the region, $\delta=(\Delta t/t_{E})^{-4}\ll 1$.

      The second set of effects is due to the fact that the time-averaged 
      distribution of the local state is not ideal. There are two sources of
      deviation: spread and bias.
\begin{description}
  \item[Spread:] The local states do not go precisely to some pointer-basis 
      state after each application of the chaotic map. Their distribution has
      a spread about each of these, which is small because of the shift map.
      In $\S 5$, we find this spread to be about the same as the exponential
      factor above, $\exp-(\Delta t/t_{E})^{2}\ll 1$. 
  \item[Bias:] The complete chaotic map for the local state has no bias, in the
      sense that its distribution is symmetric under exchange of pointer-basis
      states. While most paths are not symetrically distributed, the average
      over all possible initial states for the path {\em is} symmetric.
      The only bias is introduced by the attraction of the GDM, 
      which is {\em not} usually at a symmetric 
      point(i.e. multiple of the identity). Thus in the two-state system, the
      GDM will push the distribution toward one of the two states,
      away from the other, at any given time. This effect is very small, 
      because of the small spread of the distribution and the large mixing
      of the chaotic map. First consider the case, where the GDM
      is diagonal in the pointer basis. For any local state which is near
      a pointer state, the GDM has a negligible effect (unless
      it is inside the capture region!). It can only succesfully 
      pull, from one half of the Hilbert space to the other, those local 
      states which have comparable components of the two basis states.
      But the magnitude of the distribution away from these is suppressed 
      by the factor $\exp-(\Delta t/t_{E})^{2}$, according to $\S 5$. Hence
      this effect should be supressed by this factor. If the GDM is not
      diagonal, then it has a sizable effect on the tight distribution about
      each pointer-basis state, pulling it a considerable distance around the 
      Hilbert space. However, before the distribution is sampled by the
      global state, another cycle must be completed, during which the
      chaotic map will mix the distribution beyond recognition, and
      return it to the vicinity of the pointer-basis states. If this
      mixing is strong enough ($\Delta t/t_{E}$ big enough), the bias will be
      eliminated, or replaced with a bias which is statistically unrelated
      to the density matrix. It is difficult to estimate this effect, and
      below we will assume the GDM is diagonal, due to the strong
      decohering interactions.
\end{description}

      Now let us see what effect these deviations from ideality will have on
      the true probabilities $q(z)$. In the two-state case, the local state
      is distributed in the following way. With probability $(1-\delta f)/2$,
      it is within $\delta_{0}$ of the state coresponding to $z=0$, while
      with probability $(1+\delta f)/2$, it is within $\delta_{1}$ of the 
      state corresponding to $z=1$. This alters the average value of $z$ for
      the GDM after one cycle, or equivalently the true probability $q$. 
\[ q(t)\rightarrow \bar{q}(t+\Delta t)=q(t)+\overline{\delta q}(t),  \]
      where $\overline{\delta q}=0$ in the idealized case, 
      but here instead
\[ \overline{\delta q}=\beta q(1-q)[\delta f(1-\delta_{+})-\delta_{-}], \;\;
            \delta_{+}=\delta_{1}+\delta_{0}, \;\;
            \delta_{-}=\delta_{1}-\delta_{0}.             \]
      The number of cycles that pass, on average, before the capture
      process results in a measurement, is about $\beta^{-2}$, and the
      effect is (mostly)cumulative, rather than difussion-like. Therefore
      the expected total drift should be
\begin{eqnarray}
   \overline{\delta q}_{total}
       &\approx& q(1-q)[\frac{\delta f-\delta_{-}}{\beta}], \nonumber \\
       &\approx& e^{-(\Delta t/t_{E})^{2}}/\beta, \nonumber \\
       &\ll&\delta=(\Delta t/t_{E})^{-4}.  
\end{eqnarray}
      The conclusion is, therefore, that the dominant factor in making the 
      frequencies of measurement outcomes differ from BP is the
      finite size of the capture region $\delta$. Thus the largest deviations
      from BP will be found whenever the decohering interactions, which
      actually cause a measurement, are not far above the threshold for
      the measurement to take place, i.e. just inside the chaotic regime.

\part*{II: Applications}
\addcontentsline{toc}{part}{Part II: Applications}
\begin{quotation} \footnotesize
    Before Zen, a tree is a tree and a mountain is a mountain.
    During Zen, a tree is not a tree and a mountain is not a mountain.
    After Zen, a tree is again a tree and a mountain is again a mountain.
    {\em Chuang-Tzu}
\end{quotation}
\section{Standard Measurement Situations}
      In this section we try to connect the result of Part I to real
      measurements. First we discuss a simple detector, which registers the
      fact that its environment has reached a certain pre-selected state.
      Then the detector is generalized to select from among a number of
      possible states of a system of interest. Finally we consider the effect
      of kinetic terms in the local Hamiltonian, which lead to intermittent
      behaviour, i.e. jumping around from one pointer basis state to another.

\subsection{A simple detector}
      In the simplest measurement scenario, we call the local degrees of 
      freedom the detector, which monitors the environment for a certain
      state $\ket{a_{0}}$. The local and global states are
\begin{eqnarray*}
     {\rm local} &:&\ket{s}=\sum_{n}s_{n}\ket{n}, \;\; n=0,1    \\
     {\rm global}&:&\ket{\psi}=\sum_{na}\psi_{na}\ket{na}, 
                    \;\; n=0,1; \;\; a=1,2,\ldots ,A.
\end{eqnarray*}
      The required Hamiltonian is the sum of three parts:
\begin{eqnarray}
    \hat{H}       &=&\hat{H}_{env}+\hat{H}_{dec}+\hat{H}_{meas}  \\
    \hat{H}_{env} &=&{\rm kinetic \;energy \;of \; environment} \nonumber  \\
                  &=&\sum_{ab}K_{ab}\proj{b}{a}
                     \otimes \sum_{n}\proj{n}{n}                 \\
    \hat{H}_{dec} &=&{\rm decohering \;interactions}  \nonumber  \\
                  &=&\sum_{na}E_{na}\proj{na}{na}                \\
    \hat{H}_{meas}&=&{\rm active \;interactions}   \nonumber  \\
                  &=&B(\proj{1a_{1}}{0a_{0}}+\proj{0a_{0}}{1a_{1}}), \;\; 
                     \braket{a_{0}}{a_{1}}=0.
\end{eqnarray}
      We first show that the effective local Hamiltonian can be neglected.
      It is calculated as above, by assuming no information about the
      environment, or equivalently tracing the Hamiltonian over the
      environment degrees of freedom.
\[ \hat{H}_{local}=\sum_{mn}V_{mn}\proj{n}{m}    \]
\begin{eqnarray*}
    V_{mn}&=&\frac{1}{A}\sum_{a}\matel{ma}{\hat{H}}{na}      \\
          &=&\frac{\delta_{m,n}}{A}\sum_{a}(K_{aa}+E_{na})   \\
          &=&\delta_{m,n}(K+\frac{1}{A}\sum_{a}E_{na})       \\
          &=&\delta_{m,n}\times {\rm constant, \;by \;assumption.}
\end{eqnarray*}
      The only effect the Hamiltonian has on the local state is therefore
      the map induced by $\hat{H}_{dec}$, which is assumed to
      be in the strongly chaotic regime in order to effect a measurement. The
      presence of $\hat{H}_{meas}$ has a negligible impact on this evolution,
      provided $B$ is small compared to the decoherence energy. To see this,
      we calculate the decoherence function ${\cal Z}$ which generates the map.
      Proceeding as in $\S 4.2$, we find that it is the sum of two parts:
\begin{eqnarray}
          {\cal Z}&=&{\cal Z}_{dec}+{\cal Z}_{meas},              \\
    {\cal Z}_{dec}&=&\frac{1}{2}
                     \left(\frac{E_{dec}\Delta t}{\hbar}\right)^{2}
                     \sum_{n}p_{n}(1-p_{n}), \;\;p_{n}=\nsq{s_{n}},  \\
                  &=&{\rm the \;usual \;term},    \nonumber          \\
   {\cal Z}_{meas}&=&\frac{1}{NA}
                     \left(\frac{B\Delta t}{\hbar}\right)^{2}
                     (p_{0}^{2}+(1-p_{0})^{2})\ll 1,                  \\
   {\cal Z}_{meas}/{\cal Z}_{dec}&\sim&
                    \left(\frac{B}{E_{dec}}\right)^{2}/(NA)\ll 1.
\end{eqnarray}
      Note that it is actually the gradient, in the local Hilbert space,
      of ${\cal Z}$ which determines the map. Hence the fact that, at the
      pointer basis states, ${\cal Z}_{dec}=0$ while ${\cal Z}_{meas}\neq 0$
      is not important. So, the local state is governed by the usual chaotic
      map, which makes it jump pseudo-randomly between the pointer basis states
      $\ket{n}$.

      Let us discuss the role of the different terms in the Hamiltonian
      as they affect the global state, and then go through the measurement
      itself from beginning to end. $\hat{H}_{dec}$ is a strong decohering
      term and will force the GDM,
\[ \hat{Q}=\sum_{a}\matel{a}{(\proj{\psi}{\psi})}{a}
                 =\sum_{mna}\psi_{ma}\psi_{na}^{\ast}\proj{m}{n}\;,    \]
      to remain diagonal throughout the process. I.e.
\( \hat{Q}\rightarrow \sum_{n}p_{n}\proj{n}{n}.              \)
      $\hat{H}_{meas}$ will effect the measurement by moving the detector
      away from its set(initial) state when the environment is in the
      selected state $\ket{a_{0}}$. $\hat{H}_{env}$ has the effect of mixing
      up the environment states. This will make it possible for the
      measurement to require a finite amount of time before occurring, and for
      the outcome to persist for some finite period of time.
      Let us consider the evolution of the global state in the absence of
      the local one, and then see how this is affected by its presence.
      The state begins at
\[ \ket{\psi_{0}}=\ket{0}\otimes \sum_{a}C_{a}\ket{a}, 
                 \;\;\nsq{C_{a_{0}}}=0.                  \]
      As a result of $\hat{H}_{env}$, the state eventually moves to
\[ \ket{\psi_{0}^{\prime}}=\ket{0}\otimes 
        \sum_{a}C_{a}^{\prime}\ket{a},\;\;\nsq{C_{a_{0}}^{\prime}}>0.  \]
      Now we need to anticipate a result of $\S 8.3$: The detector decays 
      toward the triggered state via an exponential(approximately) given by
      $k=e^{-\lambda t}$, where $\lambda \sim B^2/(\hbar E_{dec})$.
      The state becomes
\[ \ket{\psi (t)}=k\ket{0}\otimes \ket{A_{0}(t)}+
                 \sqrt{1-k^{2}}\ket{1}\otimes \ket{A_{1}(t)},   \]
      where the environment states are complicated.
      Eventually(asymptotically as $t\rightarrow \infty$), the state reaches
      the triggered state: 
          $\ket{\psi_{\infty}}=\ket{1}\otimes \ket{A_{\infty}}.$
      Thus the detector will be triggered, if the environment state remains
      near the selected state $\ket{a_{0}}$ forever. More realistically, and
      for finite times, the probability of triggering will not equal one,
      and the quantum description of the detector will remain a mixture 
      of the two states. In practice, either the detector
      is triggered or it is not, with a probability which depends on the
      experimental arrangement, i.e. the initial state of the environment and
      $\hat{H}_{env}$. 
 
      Now let us see how MOM affects the evolution of the detector.
      At the beginning, the local and global states
      agree on the detector state, and the environment is not yet ready
      to trigger the detector:
\[ \ket{s}=\ket{0}, \;\; \ket{\psi}=\ket{0}\otimes 
                 \sum_{a}C_{a}\ket{a}, \;\; \nsq{C_{a_{0}}}=0.   \]
      A period of time follows during which nothing happens to the
      detector, because $\nsq{C_{a_{0}}}$ is too small. Although the active
      interactions try to move $p_{0}$ away from 1, the attraction to
      the local state keeps it there: A threshold must be exceeded by
      $\nsq{C_{a_{0}}}$ at some point, or the detector will not be triggered.

      To see this, consider the combined effect of the active
      interaction(decay) and the attraction to the local state on the GDM.
      First allow the attraction map to operate,
      recalling that we expect $\beta \sim 1/A \ll 1$:
\[ p_{0}\rightarrow p_{0}^{\prime}=p_{0}+\beta p_{0}(1-p_{0}). \]
      Then allow the decay:
\[ p_{0}^{\prime}\rightarrow p_{0}^{\prime \prime}=p_{0}^{\prime}e^{-\epsilon},
   \;\;\epsilon=\nsq{C_{a_{0}}}B^{2}\Delta t/(\hbar E_{dec}).  \]
      We want to see the total effect on the difference $\delta =1-p_{0}$.
      Neglecting small terms, and setting $\delta =\delta^{\prime \prime}$,
      we reach after some algebra the condition
      $\delta=\epsilon/\beta \equiv \delta^{\ast}$.
      For $\delta <\delta^{\ast}$, $\delta$ grows, while for
      $\delta >\delta^{\ast}$, it decreases: It is an attracting fixed point.
      Therefore $p_{0}\rightarrow 1-\delta^{\ast}$ as long as the local state
      remains stuck at $\ket{0}$. The question then arises, whether
      $\delta^{\ast}$ is small enough to be still in the capture region. 
      In Part I($\S 7.2$) we found this region to be bounded by
      $p_{0}>1-p^{\ast}, \;\; p^{\ast}\sim (\Delta E/E_{dec})^{4}$,
      where $E_{dec}$ is
      the strength of $\hat{H}_{dec}$, and $\Delta E=\hbar/\Delta t$.
      Now, if $\delta^{\ast}<p^{\ast}$, the local state will remain trapped and
      the decay will never occur, while if $\delta^{\ast}>p^{\ast}$, 
      the local state
      will soon leave $\ket{0}$ and undergo chaotic motion due to the
      effect of $\hat{H}_{dec}$. In this case, the probability $p_{0}$ will
      execute a pseudo-random walk, as described in Part I. The condition
      on $\nsq{C_{a_{0}}}$ for the local state to be released is
\begin{equation}
       \nsq{C_{a_{0}}}>\beta
       \frac{\Delta E^{5}}{B^{2}E_{dec}^{3}}\equiv P_{crit}.
\end{equation}
      There are two possibilities. Either $P_{crit}<1$ and release {\em may}
      occur, or \linebreak $P_{crit}>1$ and release {\em cannot} take place.
      This becomes a condition on $B$ for the detector to be triggered at all:
\begin{equation}
         B^{2}>\beta (\Delta E)^{2}(\Delta E/E_{dec})^{3}.
\end{equation} 
      Note that we are assuming the ratio $\Delta E/E_{dec}$ is small, or the 
      local state dynamical map would not be chaotic, and the random walk
      would not occur. Hence this condition could to be satisfied. 
       
      Once the local state has been released from $\ket{0}$
      it will undergo a chaotic dynamics, jumping between the
      two basis states, $\ket{0}$ and $\ket{1}$.
      This will cause the global state to undergo the pseudo-random 
      probability walk treated in Part I, except that it will also be
      moving toward $p_{1}=1$, with a slow exponential decay. There will be two
      competing effects: the slow decay and the fast tendency to reach
      a pointer basis state. To see how these work together, it is crucial to
      note that the probabilities $p_{n}(t)$ are {\em not}\, affected by the
      random walk. Although the detector tends to reach a pointer basis state
      quickly, it does so strictly according to the current probabilities
      $p_{n}(t)$: No `Zeno' effect occurs, which would change the expected
      time of triggering, because of the exponential form of the decay.
      The decay has that form because of the environment kinetic terms,
      which effectively dampen the motion. Once we assume the exponential
      decay form, it is easy to see that the random walk decouples from it.
      Let $p_{0}=p$ at some time $t$. According to MOM,
      during time $\Delta t$ the average value of $p_{0}$ will evolve as
\begin{eqnarray*}
     p&\rightarrow p^{\prime}&
                 =\frac{1}{2}(p+\delta p)+\frac{1}{2}(p-\delta p)  \\
      &\rightarrow p^{\prime \prime}&
                 =\frac{1}{2}e^{-\epsilon}(p+\delta p)
                 +\frac{1}{2}e^{-\epsilon}(p-\delta p)            \\
      &&=pe^{-\epsilon},
\end{eqnarray*}
      which is the same as without the random walk. Note two important facts.
      First, it is only because of the {\em exponential}\,
      form of the decay that
      this works. Second, there will be small deviations from  the standard 
      probability evolution due to two factors: the (non-exponential)
      quadratic nature of all such decays for very short times, and the
      finite size of $p^{\ast}$.
      The quadratic effects should be small, unless the decay is still
      quadratic for $\delta \sim \delta^{\ast}$, in which case the above
      analysis is invalid: We assume this is not the case.
      The other deviations should be only of order $p^{\ast}$, 
      so when the detector
      does trigger, it should do so after being exposed to the selected
      environment state $\ket{a_{0}}$ for the usual length of time.
      The triggering does not last forever,
      however, because the environment will eventually return to the
      vicinity of the state $\ket{a_{1}}$, and the detector will then
      start moving back toward $\ket{0}$, in a manner analogous to the initial
      departure from that state. This presents a picture of the detector
      switching at unpredictable intervals between the two states. However,
      we should take into account the fact that the environment has a very
      large number of states($A\rightarrow \infty$). This means
      that the recurrence time for the state $\ket{1a_{1}}$ is effectively
      infinite, so that the result of the detector triggering will be
      permanent.
      It also means that the initial waiting period for $\nsq{C_{a_{0}}}$
      to grow above $P_{crit}$ is also arbitrary.
      The final picture is that the detector sits for an
      indeterminate length of time, waiting to be triggered, then is triggered
      according to an exponential decay law during a relatively short
      period, and then rests in the triggered state indefinitely.

\subsection{Multi-state detector/selector} 
      In this measurement scenario, the local degree of freedom is still
      the detector, but in addition to the environment there is an
      `object' system from which a state will be selected. 
      The particular state is a member of the pointer basis and is not
      known in advance, but it will be chosen according to the usual
      probability rule. The local and global states are
\begin{eqnarray*}
     {\rm local} &:&\ket{s}=\sum_{n}s_{n}\ket{n},    \\
     {\rm global}&:&\ket{\psi}=\sum_{nia}\psi_{nia}\ket{nia}, 
\end{eqnarray*}
\[  n=0,1,\ldots,N; \;\; i=1,\ldots,N; \;\; a=1,2,\ldots ,A.  \]
      The required Hamiltonian is still the sum of three parts:
\begin{eqnarray}
    \hat{H}       &=&\hat{H}_{env}+\hat{H}_{dec}+\hat{H}_{meas}  \\
    \hat{H}_{env} &=&\sum_{ab}K_{ab}\proj{b}{a}
                     \otimes \sum_{ni}\proj{ni}{ni}                 \\
    \hat{H}_{dec} &=&\sum_{nia}E_{nia}\proj{nia}{nia}                \\
    \hat{H}_{meas}&=&B\sum_{i}(\proj{iia_{i}}{0ia_{0}}
                              +\proj{0ia_{0}}{iia_{i}}), \;\; 
                     \braket{a_{0}}{a_{i}}=0.
\end{eqnarray}
      As before, the effective local Hamiltonian can be neglected, as well as
      the self-energy of the object system. The story of this measurement
      process is similar to that above. The global state begins at
\[ \ket{\psi_{0}}=\ket{0}\otimes \sum_{i}D_{i}\ket{i}\otimes 
                 \sum_{a}C_{a}\ket{a}, \;\;\nsq{C_{a_{0}}}=0, \]
      and would, in the absence of the local state, evolve asymptotically to
\[ \ket{\psi_{\infty}}=\sum_{i}D_{i}\ket{ii}\otimes\ket{A_{i,\infty}}. \]
      Once the local state has been released from $\ket{0}$
      it will undergo a chaotic dynamics, jumping between the
      basis states, $\ket{0}$ and $\ket{i}$. Finally the detector,
      object and environment will reach a semi-stable state 
      $\ket{\bar{\imath}\bar{\imath}}\otimes \ket{A_{\bar{\imath}}(t)}$,
      with probability given by $\nsq{D_{\bar{\imath}}}$, as expected.

\subsection{Intermittency}
      In the idealized version of measurements used above, the result of a
      measurement is permanent. However, actual measurements do not last
      forever. Sooner or later, the outcome of any measurement will fade away.
      This is due to the presence of kinetic terms in the effective
      Hamiltonian(self-energy) which are not diagonal in the pointer basis.
      Thus, after the position of a particle is measured, the wavefunction
      of the particle in space will spread out: A new measurement of the
      position will probably yield a different value. In terms of MOM, if
      the kinetic terms are big enough the system will show intermittent
      behavior: It will jump randomly amongst the pointer basis states at
      random intervals. However, if these kinetic terms are very small,
      they will not be effective and the system will become stuck
      permanently at one of these states.

      In the simplest case we consider a two-state system with its environment:
\begin{eqnarray*}
     {\rm local} &:&\ket{s}=u\ket{u}+v\ket{v}                   \\
     {\rm global}&:&\ket{\psi}=\sum_{a}(u_{a}\ket{ua}+v_{a}\ket{va}), 
                    \;\; a=1,2,\ldots ,A.
\end{eqnarray*}
      The required Hamiltonian contains a kinetic(self-energy) part and
      a decohering part:
\begin{eqnarray}
    \hat{H} &=&\hat{K}+\hat{D}                                   \\
    \hat{K} &=&\omega (\proj{u}{v}+\proj{v}{u})
                     \otimes \sum_{a}\proj{a}{a}                 \\
    \hat{D} &=&\sum_{a}(E_{a}\proj{ua}{ua}+F_{a}\proj{va}{va}).
\end{eqnarray}
      To see the effect of these two terms, we consider the GDM, $\hat{Q}$,
      and change over to the Riemann variables $(x,y,z)$:
\begin{equation}
       \hat{Q}=\left[ \begin{array}{cc}
                        p&qe^{i\theta}  \\
                        qe^{-i\theta}&1-p
                      \end{array} \right]
              =\left[ \begin{array}{cc}
                        \frac{1}{2}(1+z)&x-iy \\
                        x+iy&\frac{1}{2}(1-z)
                      \end{array} \right].
\end{equation}
      Now we write differential equations for $(x,y,z)$. The kinetic term is
      easily handled, but the decohering interactions must be approximated.
      Their effect is to exponentially decay the off-diagonal parts of
      $\hat{Q}$: This is represented by terms proportional to $\lambda$,
      which represents the strength of these decohering interactions.
      Defining $k\equiv -2\omega$, the evolution is
\[ \frac{dx}{dt}=-\lambda x \;, \;\;
   \frac{dz}{dt}=-ky \;, \;\;
   \frac{dy}{dt}=kz-\lambda y. \]
      We let $x=0$ for simplicity. This is a linear system and can be solved
      by finding the eigenvalues and eigenvectors of the time evolution
      operator. We are interested in the overdamped case:
      $R\equiv \lambda /(2k)\gg 1.$ The eigenvalues are
\begin{eqnarray*}
      \alpha_{+}&=&-k(R-\sqrt{R^{2}-1})\approx -k^{2}/\lambda \ll k  \\
      \alpha_{-}&=&-k(R+\sqrt{R^{2}-1})\approx -\lambda.
\end{eqnarray*}
      The eigenvectors are approximately
\begin{eqnarray*}
      \vec{r}_{+}&=&\left( \begin{array}{c} z \\ y \end{array} \right) 
           =\left( \begin{array}{c} 1 \\ k/\lambda \end{array} \right)
           \approx \left( \begin{array}{c} 1 \\ 0 \end{array} \right)   \\
      \vec{r}_{-}&=&\left( \begin{array}{c} k/\lambda \\ 1 \end{array} \right)
           \approx \left( \begin{array}{c} 0 \\ 1 \end{array} \right).
\end{eqnarray*}
      Apparently, the net result is that the components $(x,y)$ decay very
      rapidly, almost exactly as they would without the kinetic term, while
      the component $z$ decays very slowly. The relationship between the
      time scales is
\begin{equation}
        T_{decay}T_{decohere}=T_{kinetic}^{2},
\end{equation}
\begin{eqnarray*}
      T_{decay}&=&\lambda /k^{2}:{\rm z-decay \;time},                     \\
   T_{decohere}&=&1/\lambda :{\rm decoherence \;or \;(x,y)-decay \;time},  \\
    T_{kinetic}&=&1/k:{\rm natural \;spreading \;time}.
\end{eqnarray*}
      Now we want to consider the effect of the kinetic term on the 
      random walk of the $z$-component of the GDM,
      and on the chaotic map of the local state.
      As long as $k\ll \Delta E$, the chaotic map will be unaffected, and
      the local state will behave overall as usual. The random walk for $z$
      will be as in $\S 7.1$, except that $\ket{u}$ and $\ket{v}$ are
      {\em not} in general absorbing states.
      By a similar analysis we can find the condition for
      the system to be able to escape $\ket{u}$ or $\ket{v}$, in
      case both the local and global states find themselves there. The kinetic
      terms must be strong enough:
\begin{equation}
     k>k_{min}=\beta^{1/2}\Delta E(\frac{\Delta E}{\lambda})^{3/2}.
\end{equation}
      With our usual assumptions ($\beta \ll 1 \;and \;\lambda \gg \Delta E$)
      this is possible, since $k_{min}\ll \Delta E$.
      Assuming this is the case, $z$ executes a random walk
      in the interval between $\pm (1-\delta^{\ast})$, where
      $\delta^{\ast}=\Delta t/(\beta T_{decay})$. This is the combination of
      the basic random walk due to the attraction to the local state, and the
      decay which makes $z\rightarrow 0.$ We can see what the overall
      pattern of behavior will be by noting that the motion is very similar
      to a {\em random walk with reflecting barriers}\cite{Feller}.
      Near the center, for $z\approx 0$, the decay is unimportant and the 
      random walk is unbiased. At the edges, for $z=\pm (1-\delta^{\ast})$,
      $z$ either stays unchanged or moves one step toward the center, 
      with both probabilities equal to $1/2$. In between, there is a smooth
      transition from one type of behavior to the other. Assuming the 
      analogy is valid, we can use Feller's results. Let us start $z$ at an
      arbitrary state, and consider the distribution of values it will
      have after some large number of time steps. For an unbiased random
      walk with reflecting barriers and steps of the same size everywhere,
      this tends toward a uniform distribution. In our case the steps are 
      not of the same size, rather they are roughly 
      $\Delta z\approx \beta (1-z^{2})$. This means the distribution is
      concentrated where the steps are smallest, where $1-z^{2}$ is small,
      but diminishes to zero near $z=\pm 1$. The effect of the $z$-decay,
      because it introduces a bias toward the center, will be to soften
      this distribution somewhat. If this is not overwhelming, we should see
      the following behavior. The system (global $z$) spends almost all of its
      time near the states \mbox{$z=\pm 1$,}
      with ocassional departures into the 
      center: These last only for a short time, about $\Delta t/\beta^{2}$.
      If we start with the system at, e.g. $z=1$, and ask what is the
      probability for the system to be there again at a later
      time, the answer is very nearly the standard exponential
      decay towards 1/2, which we expected in the absence of the local
      state introduced by MOM. Again, this is because the exponential
      decay `commutes' with the random walk. Note that in the standard
      interpretation of the behavior of the global state alone, without
      the added elements of MOM, this answer corresponds to a different
      question, namely , if we measure the system to be at $z=1$, what is the
      probability that a subsequent measurement will give the same result?
      Within MOM description of the situation, no intervention by an 
      outside observer is needed to explain the odd behavior of the system:
      It behaves this way already without measuring apparatus at hand.
      There are several experiments which exhibit this kind of 
      intermittent behavior \cite{atom,SQUID}.
      The standard explanation for the behavior is that it is {\em as if}
      the state of the system, along the pointer basis, were measured
      periodically, resulting in an exponential decay law via the Zeno effect.
      The actual exponential decay rate depends on the frequency of these
      measurements, which is taken to be given by the decoherence time
      \cite{Leg}. In MOM, there are no special assumptions needed to obtain
      the intermittency. I.e. this follows from the same general model
      which explains normal measurements. (See also $\S 13$).

\section{Non-standard Situations}
      In this section we consider the possible effect of applying MOM
      to certain situations which are not like those of $\S 8$, but may
      correspond to observable behavior.
      In each case, we assume the domain inside the cut contains only
      a specific degree of freedom, which interacts with its environment
      in a specified way. Then we calculate the decoherence function
      ${\cal Z}$, from which we can anticipate qualitative aspects of the
      evolution of the combined local and global system.

\subsection{Spin observer w. spin-vector interactions}
      One may take the local degrees of freedom to be the spin of some system,
      a field or particle, which interacts via a spin-vector dot product
      coupling with the environment. This environment can be a set of
      particles with spin, a vector or tensor field, or any system which
      allows a representation of a vector operator. 
      We begin with the simplest case:
\[ {\rm local}:\ket{s}=\sum_{m}s_{m}\ket{m}, \;\;
         {\rm global}:\ket{\psi}=\sum_{mm_{e}}\psi_{mm_{e}}\ket{mm_{e}},  \]
\[ m=-j,-j+1,\ldots ,+j; \;\; m_{e}=-j_{e},-j_{e}+1,\ldots , +j_{e}. \]  
      The Hamiltonian is a spin-spin interaction with strength $E$:
\begin{equation}
        \hat{H}=(E/\hbar^{2})\vec{J}\cdot \vec{J}_{e}\;.
\end{equation}
      The effective Hamiltonian for the local state is zero, and the
      decoherence generating function is
\begin{eqnarray}
   {\cal Z}&=&(1/3)(\frac{E\Delta t}{\hbar})^{2}j_{e}(j_{e}+1)
              (\matel{s}{\vec{J}^{2}}{s}-\nsq{\matel{s}{\vec{J}}{s}})
              /\hbar^{2}                        \nonumber             \\
           &=&(1/3)(\frac{E\Delta t}{\hbar})^{2}j_{e}(j_{e}+1)
              (j(j+1)-\nsq{\vec{J}}/\hbar^{2}) \;,
\end{eqnarray}
      where $\nsq{\vec{J}}\equiv \nsq{\matel{s}{\vec{J}}{s}}$.
      Much can be determined from examination of this function. 
      First note that the effective strength of the interaction is actually
      $E_{eff}=Ej_{e}$, which can be much greater than $E$. 
      Now let us consider some simple examples.
      For $j=0$ there is {\em no} local degree of freedom. For $j=1/2$, \\
      ${\cal Z}$ is constant and there
      is no map induced on the local state: A spin-1/2 system local state
      is immune to the decoherence effects. As we will see below,
      this is because it already has its spin maximally aligned along 
      some direction. Now let $j$ be arbitrary, but restrict the state by 
      $\ket{s}=\ket{m}$. Then $\nsq{\vec{J}}=m^{2}$, and ${\cal Z}$ has
      a simple dependence on $m$.
      Assuming $j$ to be an integer, we calculate the ratio $R$ of ${\cal Z}$
      for $m=0$ to that for $m=j$: $R=j+1$. This means
      that the gradient of ${\cal Z}$ becomes greater as $j$ is increased.
      As we see below, this leads toward more classical
      behavior, for a given value of $E$.

      Next we want to find the fixed points of the mapping induced 
      on the local state. This map is proportional to the gradient of
      ${\cal Z}$, so these are its maxima and minima.
\begin{description}
  \item[maxima:] These are the points with $\nsq{\vec{J}}$ as small as
    possible,
    which are linear combinations of the states $\ket{m}$,
    with amplitudes of equal norm. These states are much like plane-waves
    or standing waves in mechanics. For very small values of 
    $E_{eff}$, the shift map can be ignored. The map slowly pushes
    the local state toward one of these states, the closest
    one to the initial state. This will pull the 
    global state also toward this kind of state, or perhaps toward an 
    equal-probability mixture of these states. Note that this is decidedly
    {\em non-classical} behavior, because classical spins always have
    maximal value of $\nsq{\vec{J}}$.
  \item[minima:] These are the points with $\nsq{\vec{J}}$ as large as
    possible,
    namely all states where $\matel{s}{\vec{J}}{s}=\hbar j\hat{e}$,
    where $\hat{e}$ is a unit vector in any direction. For large values of
    $E_{eff}$, the local state will be in the chaotic regime, and will
    be driven at the end of each cycle $\Delta t$ to one
    of these states. As a result, the global state is always being pulled
    toward a {\em classical} state, i.e. one with definite spin direction.
\end{description}
            
      Now we consider the gradient of ${\cal Z}$, which governs the local-state
      behavior as follows. If $E_{eff}$ is small, the induced map is given
      just by the discrete map, and the local state tends toward the maxima
      of  ${\cal Z}$, where the gradient is zero. If $E_{eff}$ is large,
      the discrete map is chaotic, and the shift map becomes
      dominant. Recall that this map is the result of evolving
      the state $\ket{s}$ for time $\Delta t$ under the equation
\[ d\ket{s}/dt=-\ket{\vec{\nabla}_{\bot}{\cal Z}}.  \]
      Therefore the final state after each cycle will tend to be
      near some minimum of the gradient, which is given by
\[ \ket{\vec{\nabla}_{\bot}{\cal Z}}=(k/\hbar^{2})\hat{G}\ket{s},  \]
\begin{equation}
      \hat{G}=-(1-\proj{s}{s})\matel{s}{\vec{J}}{s}\cdot \vec{J} \;,\;\;
            k=(2/3)(\frac{E\Delta t}{\hbar})^{2}j_{e}(j_{e}+1).
\end{equation}
      This equals zero only at the points 
      $s_{m}=\delta_{m\bar{m}},\;\bar{m}=-j,-j+1,\ldots ,+j.$
      Defining $\omega$ and $p_{m}$:
\[ \matel{s}{\vec{J}}{s}\equiv\hbar\omega\hat{z},
                        \;\; p_{m}\equiv \nsq{s_{m}},  \]
      the shift map is governed by 
\begin{equation}
          dp_{m}/dt=k\omega(m-\omega)p_{m} \;,
\end{equation}
      which in particular implies
\[ dp_{j}/dt\geq 0 \;,\;=0\;\;\;
         {\rm iff}\;\;\;\omega=j:{\rm classical\;state}. \]
      This supports the statements made above concerning the qualitative 
      behavior of the local state.

      Now let us extend the local system and environment by allowing
      different values for the spin magnitudes:
\[ {\rm local}:\ket{s}=\sum_{jm}s_{jm}\ket{jm}, \;\; {\rm global}:\ket{\psi}=
   \sum_{jmj_{e}m_{e}}\psi_{jmj_{e}m_{e}} \ket{jmj_{e}m_{e}},   \]
\[ j=0,1,2,\ldots,J; \;\; j_{e}=0,1,2,\ldots,J_{e}.  \]
      Up to numerical factors of order one, ${\cal Z}$ does not change,
      except that $j_{e}$ is replaced by $J_{e}$. The gradient changes
      because of the expansion of the local Hilbert space, such that
      the new $\hat{G}$ is given by
\begin{equation}
   \hat{G}=-(1-\proj{s}{s})(2\vec{J}-\matel{s}{\vec{J}}{s}) \cdot \vec{J} \;.
\end{equation}
      The result, for large $E_{eff}$, is that the local state, 
      in addition to tending toward states with maximal spin projection
      along some axis as before, now also tends toward states with
      small values of $j$. In particular, $j=0$ is favored above all others.
      To see this, define $p_{j}\equiv \sum_{m=-j}^{+j}\nsq{s_{jm}}$. 
      The shift map is such that
\[ dp_{j}/dt=0 \;{\rm when} \;\; p_{j}=\delta_{j\bar{\jmath}} \;, 
                 \;\; \bar{\jmath}=0,1,2,\ldots,J \;,  \]
      and
\[ dp_{0}/dt\geq 0\;,\;=0\;\;{\rm iff}\;\;p_{0}=0\;{\rm or}\;1 \;.  \]
      Thus the local state is drawn inexorably toward $j=0$. The only escape
      is the possibility that this tendency will be ameliorated by the
      attraction to the global state. Indeed, it is not likely that
      the global state can actually go to and {\em stay} at $j=0$,
      because of other interactions which we have ignored. For example,
      if the local system spin is associated with a particle, this
      particle's position(and spin) will interact with other particles'
      positions(and spins): This will likely keep the spin away from zero.
      Absent a detailed analysis, we can guess that the actual behavior
      of the spin system would be that it will tend to maintain a
      relatively low value(thermally determined?) for its average value
      of $j$, and that its
      spin will be maximally aligned along some direction. This direction is
      arbitrary, and would change due to collisions and other interactions.

      In summary, the qualitative behavior of the local spin is as follows.
      If the system is $j=0$ or $1/2$, then it is
      unaffected. For any other system with fixed $j$,
      the dynamics depends on whether 
      $E_{eff}$ is small or large. If it is small, then the state is driven
      toward a uniform state. If it is large, it acts to force the spin to
      have maximal projection along some spatial axis. The actual direction
      is arbitrary and will depend on the initial conditions. Note that this
      is the behaviour of classical spinning systems. If $j$ is not fixed,
      but can take any values, then the dynamics has the
      same effect, but it also forces the length of the spin to take the
      minimum value possible. This will be either $j=0$ or $1/2$,
      but other interactions(which were neglected above) may force this 
      value to rise. Note that there should be no intermittency effect,
      as such, for spin observables,
      because the relevant kinetic terms are absent from the Hamiltonian.

\subsection{Position observer w. Coulomb-type interaction}
      One may define the local degree of freedom to be the set of positions
      of some particles, which interact with other particles or a
      matter-density field
      through a Coulomb-type gravitational term. The simplest case is where
      the local system and its environment consist of just one particle each.
      For convenience, let us use a discretized version of three-dimensional
      space, with a finite number $N$ of sites, each of small size $R_{1}$.
      If the size of the universe is $R$, we have approximately 
      $N=(R/R_{1})^{3}$. The states are
\[ {\rm local}:\ket{\psi}=\sum_{a}\psi_{a}\ket{a},\;\;
                 {\rm global}:\ket{\Psi}=\sum_{au}\Psi_{au}\ket{au},   \]
\[ a,u=1,2,3,\ldots,N. \] 
      The decoherence function depends only on a Coulomb-type interaction 
      representing gravity. If $D_{ab}$ is the positive distance between two
      sites($D_{aa}\equiv R_{1}$), then the potential is
      $V_{ab}=k/D_{ab}$, where 
      $k=Gmm'$, $G$ is Newton's constant, $m$ is the mass of the local-state 
      particle, and $m'$ is the mass of the environment particle.
      Then the (interaction) Hamiltonian is
\begin{equation}
         \hat{H}^{int}=\sum_{au}\proj{au}{au}V_{au} \;.
\end{equation}
      The contribution from this interaction to the effective Hamiltonian
      for the local state is a constant:
\[ \hat{H}_{eff}^{int}=-M\sum_{a}\proj{a}{a} \;,\;\;
                 M=(k/N)\sum_{b}D_{ab}^{-1}:\mbox{independent of a.}  \]
      The decoherence generating function ${\cal Z}$ is
\begin{equation}
  {\cal Z}=(\Delta t/\hbar)^{2}(\sum_{a}p_{a}K-\sum_{ab}p_{a}p_{b}K_{ab}) \;,
\end{equation}
      where $p_{a}\equiv \nsq{\psi_{a}}$, and
\[ K=(k^{2}/N)\sum_{b}D_{ab}^{-2}:\mbox{independent of a,}  \]
\[ K_{ab}=(k^{2}/N)\sum_{c}D_{ac}^{-1}D_{bc}^{-1}:
             \mbox{dependent only on} \;D_{ab} \;.   \]

      Before proceeding further, it is instructive to look at the simplest
      examples. 
\begin{description}
  \item[localized state:] If $\psi_{a}=\delta_{ab}$, then ${\cal Z}=0$.
          I.e. the lowest possible value for ${\cal Z}$ is realized by any
          wavefunction which is completely localized. In fact, these are the
          {\em only} states for which ${\cal Z}=0$, barring accidental 
          degeneracies due to space curvature. In the strongly chaotic regime,
          when the local-state map is dominated by the shift map,
          it nearly always returns to one of these states after each cycle
          $\Delta t$. The local state will jump around between
          these localized states, and the global state will perform a
          probability walk as in $\S 7.1$. Thus we should
          see a tendency for the particle's position to become well-defined,
          which will be partially offset by the effect of kinetic energy terms.
  \item[two-site state:] If 
          $\psi_{a}=\sqrt{p}\delta_{ab}+\sqrt{1-p}\delta_{ac}$, for $b\neq c$,
          then ${\cal Z}\propto p(1-p)$. This is just like the generic case
          of $\S 7.1$. When the 
          strength of the interaction is weak, the local state moves slowly 
          toward uniformly-distributed states(plane-waves and standing-waves).
          When the interaction is strong enough, it jumps randomly between
          the two localized states. Taking the kinetic energy terms into
          account, the particle will either: a)not be affected at all, for
          extremely weak interaction, b)become a good-momentum state,
          for moderately weak interaction, or c)become intermittently
          localized, for strong interaction.
\end{description}

      To obtain the general form of ${\cal Z}$ and judge the strength of the
      interaction, we replace the finite sums with continuum integrals
      and evaluate them using a crude regularization procedure, where all
      space integrals are cut off at the distance $R$. We obtain
\begin{eqnarray*}
       M&=&(3/2)k/R,                    \\
       K&=&3(k/R)^{2}=(4/3)M^{2},       \\
  K_{ab}&=&(3/2)(k/R)^{2}(2-D_{ab}/R)=(2/3)M^{2}(2-D_{ab}/R).
\end{eqnarray*}
      From this it follows that
\begin{equation}
     {\cal Z}=(2/3)(\frac{M\Delta t}{\hbar})^{2}\bar{D}/R \;, \;\;
                 \bar{D}=\sum_{ab}p_{a}p_{b}D_{ab},
\end{equation}
      where $\bar{D}$ will be referred to as the {\em spread}.
      The need for regularization makes it difficult to tell whether
      this is large or small, compared to one, in any particular case.
      To solve this problem, let us assume that all constants appearing
      are to be replaced by Planck quantities, except for the particle 
      mass $m$, and try to justify this replacement {\em a posteriori}. 
      I.e., let
\begin{equation}
    {\cal Z}\rightarrow (m/M_{p})^{2}\bar{D}/R_{p} \;,
\end{equation}
      where(speed of light $c=1$)
\begin{eqnarray*}
   M_{p}&=&1.2\times 10^{28}eV:Planck \;mass,  \\
   R_{p}&=&1.6\times 10^{-36}m:Planck \;length.
\end{eqnarray*}
      This will be true if
      $M^{2}/R=m^{2}/R_{p}$, which is equivalent to
      $1=(Gm'/R)\sqrt{R_{p}/R}$, or $m'=M_{p}\sqrt{N}$.

      Now we consider the following alternate situation: Let the environment
      consist, instead, of $N'$ particles of mass $m'$, each interacting with
      the local system particle in the same way as above. 
      Then ${\cal Z}\rightarrow N'{\cal Z}$, which is the same as before,
      except that $m'\rightarrow m'\sqrt{N'}$. We reach the desired result by
      letting $m'\equiv M_{p}$, and $N'\equiv N$: The particle of interest is
      interacting with a {\em sea} of particles, each with mass $M_{p}$, 
      of which there is one for each Planck-sized volume in the universe.
      This may be hard to accept at face value, but we can imagine that these
      are somehow {\em virtual} particles, associated with some deep aspect of
      sub-Planck-scale spacetime, and otherwise invisible to most
      probes. Perhaps they represent the vacuum energy of some field.

      At any rate, this development leads to interesting results.
      We can ask ourselves, at what length scale does the localization of
      the particle become dominant. I.e., for a given mass $m$, what spread
      $D$ will give ${\cal Z}\approx 1$. Putting in the appropriate constants,
      we can evaluate $D$ for any mass:
\begin{description}
  \item[electron:] $m=5\times 10^{5}eV\rightarrow D=10^{9}m$. This is
                   certainly beyond any present experimental bound.
  \item[proton:] $m=1GeV\rightarrow D=200m$. This is still large, but
                 perhaps small enough to be interesting.
  \item[heavy ion/atom:] $m=100GeV\rightarrow D=2cm$. This could be
                     experimentally accessible.
 \item[bio-molecule:] $m=10^{6}GeV\rightarrow D=2\times 10^{-10}m=2\mbox{\AA}$.
                       This is very small, but might be difficult to rule out
                  because of the difficulty in handling such large molecules.
  \item[Planck-mass] $m=M_{p}\rightarrow D=R_{p}$. Such a heavy
                     particle would remain localized at or below the
                     Planck-length. If this is an elementary
                     particle, neither MOM
                     nor any established physics can describe it: 
                    It makes no sense for a particle to have such a large mass.
                     A composite object could easily have such
                     a large mass, and then, {\em if there were a local state
                     associated with its center of mass}, its position
                     would practically become a {\em classical} object.
\end{description} 

      To summarize, the effect of an associated local state on the dynamics
      of a particle depends on the relation of the particle's spread to
      the characteristic distance $D$, which is inversely proportional to
      its mass. The relevant quantity is the spread of the global state,
      because the local state is strongly attracted to it, and will share
      its overall shape, i.e. most amplitude ratios. If the spread is
      extremely small, there is no effect and the particle behaves normally.
      If the spread is small but approaching $D$, the particle state tends
      towards approximate plane-waves of size less than $D$. Finally, if
      the spread is large compared to $D$, the particle undergoes spontaneous
      intermittent localization, as the kinetic energy of the particle
      competes with the localization effect. On a given distance scale, it
      is possible for light particles to be unaffected by the local state,
      but for heavy particles to have their non-classical spread severely
      restricted. The magnitude of the kinetic terms in the Hamiltonian
      is greater for particles with small mass and lesser for heavier ones.
      Hence the intermittency would be very slow, and possibly
      stop altogether, for very massive particles.

\subsection{Bose-Einstein and Fermi-Dirac statistics}
      Here we want to consider the evolution of a number of states, all in
      the same Hilbert space, when the Hamiltonian is zero. Only the universal
      attraction between states will be operating. The simplest case is when 
      there are two states, $\ket{U}$ and $\ket{V}$. They will coalesce
      asymptotically to the linear combination 
      $\ket{t\rightarrow \infty}\propto (\ket{U}+\ket{V})$,
      and this is monitored by following the evolution of the overlap
      probability, $z\equiv \nsq{\braket{U}{V}}$:
\begin{eqnarray*}
  \ket{U}\rightarrow \ket{U'}&=&(A\hat{P}_{V}+B(1-\hat{P}_{V}))\ket{U}\;, \\
  \ket{V}\rightarrow \ket{V'}&=&(A\hat{P}_{U}+B(1-\hat{P}_{U}))\ket{V}\;, \\
              z\rightarrow z'&=&R^{2}z \;,
\end{eqnarray*}
      where
\[ \hat{P}_{U}=\proj{U}{U} \;,\;\;
           \hat{P}_{V}=\proj{V}{V}:\mbox{projection operators,}   \]
\[ A=\sqrt{1+\beta(1-z)} \;,\;\; B=\sqrt{1-\beta z} \;,  \]
\[ R=(A-B)^{2}z+2B(A-B)+B^{2} \;.   \]
      First we note that, if the two states are identical or orthogonal,
      they remain so. Now we want to look at nearby cases:
\begin{description}
  \item[$z\simeq 0$:] 
    The states are nearly orthogonal. $\;z'=(1+\lambda(\beta))z$\,,
    \begin{eqnarray*}
      \lambda(\beta)&=&4\sqrt{1+\beta}(\sqrt{1+\beta}-1)>0\;,
                                           \;{\rm for}\;\beta>0\;,  \\
      &\stackrel{\beta \rightarrow 0}{\longrightarrow}&0\;,            \\
    &\stackrel{\beta \rightarrow 1}{\longrightarrow}&4\sqrt{2}(\sqrt{2}-1)\;,\\
      &\stackrel{\beta \rightarrow \infty}{\longrightarrow}&\infty\;.  
    \end{eqnarray*}
    The larger the value of $\beta$, the faster the two states begin to
    approach each other.
  \item[$z\simeq 1$:] 
    The states are nearly identical. Let $x\equiv 1-z$, then
    $x'=(1+\lambda(-\beta))x$,
    \begin{eqnarray*}
      \lambda(-\beta)&<&0, \;{\rm for}\;\beta>0,        \\
      &\stackrel{\beta \rightarrow 0}{\longrightarrow}&0,  \\
      &\stackrel{\beta \rightarrow 1}{\longrightarrow}&0,  \\
      &=&-1, \;{\rm for}\;\beta=3/4.
    \end{eqnarray*}
    The choice $\beta=3/4$ yields the fastest approach to $z=1$, analogous
    to critical damping of an oscillator. Apparently for larger $\beta$,
    the attraction is too large, and the states `overshoot' past each
    other(underdamping). The values $\beta=1$(zero damping),
    and $\beta=0$(infinite damping), yield the slowest possible approach,
    for small $x$. It is not clear whether this feature of the attraction
    is a real effect, or an artifact of the discrete-time `approximation.'
\end{description}
      It is noteworthy, that the evolution of $z$ for nearly-orthogonal states
      is the same as that of $x$ for nearly-identical states, except with 
      $\beta\rightarrow -\beta$. This suggests the following. If we were to 
      allow negative values($\beta=-1$) for some Hilbert 
      spaces(domains), while retaining a positive value($\beta=+1$) for others,
      the difference in behavior would be very much like the difference
      between {\em fermions} and {\em bosons}. For positive $\beta$(bosons),
      the states are attracted to each other. They will begin to coalesce
      quickly, but critically slow down as they approach a common 
      state(for $\beta=1$). For negative $\beta$(fermions), the states 
      repulse each other. They will be quickly pushed apart, toward mutually 
      orthogonal states, but will also critically slow down at the 
      end(again, for $\beta=-1$). In this case, however, it is clear that if 
      there are more states than basis states, they will be frustrated,
      and will become distributed in some other way, as far apart from each 
      other as possible. While the well-known spin-statistics
      connection in particle physics is not a prediction of MOM, this
      behavior is highly suggestive. 

      The question of invariance under time reversal is also relevant here. 
      In the discrete-time version of the 
      attraction/repulsion interaction, changing the sign of $\beta$ is 
      {\em not} equivalent to time reversal, except in the limit as 
      $\beta\rightarrow 0$. However, this limit is the same as the
      continuous-time version, which gives
\begin{equation}
    \frac{dz}{dt}=2\alpha z(1-z) \;, \;\;
    \alpha=\lim_{\beta\rightarrow 0,\Delta t\rightarrow 0}\beta/\Delta t \;.
\end{equation}
      In this case we might be able to say, that
      {\em reversing time} is equivalent to {\em exchanging bosonic and
      fermionic properties of particles}, noting that this is not to be 
      taken as special-relativistic Minkowski time, but as absolute
      Newtonean time.

      In the case of many states(more than two) in the same Hilbert space,
      it is difficult to make exact predictions beyond the above. The object of
      interest is the collective, effective density matrix:
\begin{equation}
        \hat{D}=(1/N)\sum_{i=1}^{N}\proj{\psi_{i}}{\psi_{i}} \;.
\end{equation}
      The actual form of the attraction/repulsion interaction, 
      in the case of more than two states, is not uniquely determined so far.
      Nevertheless, we can deduce some of the 
      features of the evolution of $\hat{D}$, with some plausible assumptions.
      If the attraction is acomplished pair-wise, then it is possible
      for any $\hat{D}$ to be constant in time, as long as the
      states $\ket{\psi_{i}}$ are all pair-wise either orthogonal or identical.
      If the attraction is acomplished by $\hat{D}$ acting on each state,
      then $\hat{D}$ will be constant if either a)all the states are
      eigenstates of $\hat{D}$, or b)in the diagonal basis, all the elements
      of $\hat{D}$ are either zero(say, $A-n$ of them, where $A$ is the 
      dimension of the Hilbert space) or a fixed constant equal to $1/n$.
      Finally, in any version, if $\hat{D}$ is a pure state, it will remain
      so forever. The stability of these points depends on whether we have
      attraction or repulsion. It also depends on the exact distribution of
      states in most cases, as the motion is not `autonomous.' A pure-state
      $\hat{D}$ will be stable in the attracting case, unstable in the
      repulsing case. In general, a partly or totally trivial $\hat{D}$, as
      in case b) above, will be unstable in the attracting case, stable in
      the repulsing case. If we make the boson-fermion/attraction-repulsion
      analogy, this means bosons tend toward all being in a common state,
      while fermions tend toward all being in orthogonal states. This is 
      consistent with the usual behavior of bosons and fermions, but a more
      definite prescription would be needed to take the correspondence further.

\part*{Conclusion}
\addcontentsline{toc}{part}{Conclusion}
\begin{quotation} \footnotesize
   \ldots One is tempted to suspect that the authors do not understand the Bohr
   philosophy sufficiently to find it helpful. Einstein himself had great
   difficulty in reaching a sharp formulation of Bohr's meaning. What hope
   then for the rest of us?\ldots
   While the founding fathers agonized on the question
   `particle' {\em or}\/ `wave'
   deBroglie in 1925 proposed the obvious answer
   `particle' {\em and}\/ `wave.'
   {\em J. S. Bell} \cite{Bell1}

   If we have to go on with these damned quantum jumps, then I'm sorry that
   I ever got involved. {\em E. Schroedinger} \cite{Bell3}

   Our actual situation in science is that we {\em do} use the classical
   concepts for the description of the experiments, and it was the problem of
   quantum theory to find theoretical interpretation of the experiments on
   this basis. There is no use in discussing what could be done if we were
   other beings than we are. {\em W. Heisenberg} \cite[p.56]{Heisenberg}

   God may be subtle, but he's not malicious\ldots
   God does not play dice. {\em Albert Einstein}
\end{quotation}
\begin{verse} \footnotesize
               All things near and far\\
               hiddenly linked are.\\
               Thou canst not stir a flower\\
               without the troubling of a star.\\
               \hspace*{1.25in}{\em William Blake}
\end{verse}
\section{General Principles}
    Let us consider MOM with some general physical principles in mind,
    beginning with the notion of {\bf observability.} Every observable is
    a Hermitian operator, but are all Hermitian operators observable?
    A `yes' answer reduces, but does not eliminate, the need to explain the
    measurement process: A `no' answer begs for a mechanism of some kind to
    explain the distinction. For example, Bell's {\bf beables}\cite{B4} are
    the variables in quantum field theory given objective reality, in the
    spirit of DeBroglie-Bohm. In MOM also, not all operators have definite
    values. The measurement process requires
    {\bf environmental decoherence}\cite{red,Cini,Albrecht}.
    The interactions between a local system and its environment can be so
    entangling that no pure quantum state can be ascribed to the local
    system, but only a reduced density matrix which loses coherence rapidly.
    It is often taken to represent a probabilistic mixture of orthogonal,
    observable states. By themselves, these interactions which
    correlate system and environment are not {\em sufficient}, without new
    dynamics, to precipitate measurement outcomes, but they are {\em necesary}
    within MOM. In addition, they must exceed a certain energy scale defined
    by the cyclic time $\Delta t$, and they must cross some observer domain
    cut. Not only do they decohere the global state, they also define the
    chaotic dynamics of the {\em pure} local state. The {\bf pointer basis}
    or decohering basis is important because it defines the observable to be
    measured. It is determined, if it exists, by the choice of domains for
    observers and by the full Hamiltonian. Are {\bf state-vector collapse}
    and {\bf quantum jumps} real? MOM was designed to produce collapse of the
    state vector. Allowing for intermittency effects, the system does jump
    from one classically-allowed state to another, hence these are actual
    physical phenomena.

    Another key issue is {\bf predictability}. Given the statistical nature
    of the quantum theory, it should not be easy to predict the outcome of a
    single measurement. Hence the model must be {\bf chaotic}. The dynamics of
    MOM make it very hard or impossible to predict individual trial outcomes.
    The required information lies in the fine details of the initial state(s),
    only approximately known. {\bf Uncertainty relations} are important
    predictions, and explanatory devices of the quantum theory, so must be
    respected. Except for time-energy, all the usual uncertainty relations
    are satisfied by MOM because the world is described using a Hilbert-space
    formalism. Time-energy itself is satisfied by the specific dynamics of
    MOM. In discussion of quantum measurements, the language of
    {\bf counter-factuals}\cite{Stapp} is often used. Stapp's paper offers an
    `alternative approach'(p.14) assuming that a definite outcome to a
    measurement procedure exists {\em if\/} it is carried out: If another
    observable is measured, no counter-factual assumption about a definite
    outcome need be made. (I.e. What would be the result {\em if\/} the
    original measurement {\em had\/} been made.) This applies even if the two
    observables commute. MOM obeys Stapp's assumption because every distinct
    measurement procedure involves its own Hamiltonian and unique chaotic
    evolution, even when commuting operators(observables) are involved.
    This enables MOM to avoid various EPR-type {\bf no-go theorems.}

    There are other issues of interest.
    {\bf Statistics:} The exchange symmetry of identical particles may be
    violated by a naive version of MOM. Extension to quantum fields, defined
    over finite regions of space, should solve this problem.
    {\bf Relativity:} MOM assumes the existence of a universal cyclic time
    interval governing all quantum evolution. This is Newtonean, rather than
    relativistic time, and might be defined by the rest frame of the cosmic
    background radiation.
    {\bf Non-locality:} In MOM, the global state vector maintains non-local
    correlations until a measurement outcome is selected. If this state is
    absent, however, correlations might fail over sufficiently large
    distances.
    {\bf Time asymmetry:} \cite{tasym} Zeh's paper and others suggest that
    the quantum measurement process is the origin of time asymmetry in
    physics. SE itself is time-symmetric, so one expects it cannot decribe
    all aspects of the process. MOM incorporates time asymmetry directly via
    chaotic dynamical maps, which are non-invertible: No continuous-time
    dynamics can have this property. Time reversal is explicitly broken only
    in the chaotic regime, i.e. during a measurement. Between measurements,
    the evolution is time-symmetric, closely approximating SE.

\section{Specific Models}
    Here different models and viewpoints are discussed,
    with emphasis on similarities and differences from MOM.
\subsection*{Orthodox viewpoints({\em `The tried and true.'})}
    These are not really models of measurement, but rather explanations and
    prescriptions rooted in the early triumphs of the theory, namely the
    probabilistic rule(BP) connecting theoretical quantities to experiment.
    They are the most widely accepted due to their apparent validity and
    complete lack of speculative content.

{\bf The Copenhagen School(Bohr):}
    This school of thought is well-known. The necesity to describe the
    measuring apparatus in classical language is transformed, in MOM,
    into the assumption of the existence of the cut, that certain
    systems(observers) have their own state-vectors, in addition to the
    global quantum state. The {\em unanalizability} of the measurement
    process is denied by MOM: Instead definite dynamics are introduced.
    The fundamental role of system-observer interactions in the measurement
    process is reasserted.
       
{\bf Statistical interpretation(stat):}
    This is probably the most widely-held view of quantum theory. The quantum
    state {\em does not} represent a single microscopic system, rather an
    ensemble. The probabilities defined by the Born rule are not very
    different from those defined classically, but {\em something} about small
    systems prevents a more precise description, available for larger,
    classical systems. In MOM also the probabilities are statistical, but
    there is additional information, extra quantum states which determine
    individual measurement outcomes.

{\bf Decohering histories(Dec):}
    In this picture, the quantum state of the universe is interpreted as the
    {\em incoherent} sum of a set of `histories' of the world. Each history
    is a set of possible measurement outcomes, for observables at different
    times. These can be time-dependent, and they answer the question, what
    observables have well-defined values, and when? The set of decohering
    histories may not exist, as there is a set of logical requirements they
    must satisfy, namely that the quantum state can be treated {\em as if}
    the universe were in a {\em classical} probabilistic mixture of these
    histories. The question remains, whether in reality only one of the
    histories is factual, or instead they all somehow exist in parallel.
    MOM singles out a {\em unique} result to every measurement.

\subsection*{Reactionary models({\em `We've gone too far...'})}
    Some, fearing that the quantum theory introduces too much uncertainty
    into nature, seek to re-solidify it by asserting the definite existence of
    {\em something}. The most basic instinct is to give reality to classical
    variables, but this approach can be much more subtle.

{\bf DeBroglie-Bohm model(DB):}\cite{pilot,hv,hv2}
    They postulate that in addition to the wavefunction there are
    classical variables which evolve deterministically or stochastically,
    giving objective reality to the world. The quantum state is not affected
    by them, but evolves by SE, while the classical variables respect
    the quantum probability distribution. MOM also adds
    information(extra quantum states) which helps to define a more objective
    reality for some variables, but in an indirect way: The interaction
    between quantum states induces measurements at well-defined times. In MOM,
    however, objective reality is ascribed to quantum states, not
    classical variables.

{\bf Many-worlds(MW):}
    In this picture of reality, there is a universe for every sequence of
    possible outcomes, for all measurements ever performed. Each universe
    embodies a unique outcome for each measurement, and does not interact
    with the other, `parallel' universes. This raises a question about the
    {\em role} of the observer: Does it split up into many observers after
    each measurement? The meaning of probability becomes less clear. The MW
    viewpoint dispenses with collapse, but postulates innummerable universes,
    each equally {\em real}, logically separated from each other. Also, the
    observables and splitting times are not specified. MOM is quite different,
    as it seeks to describe a {\em single} universe with {\em many} observers,
    rather than {\em many} universes each with a {\em single} observer.

{\bf Modal interpretation(mod):}
    This approach hopes, that by analysis of the quantum state, including all
    the environment degrees of freedom, it will be possible to prove, that the
    SE leads to collapse of the wavefunction. Thus, the state makes a choice
    among the possible outcomes, based on information contained in the
    environment. The observables of the theory should arise
    {\em spontaneously} from such behavior, and BP should be recovered.
    There is no evidence for this, but hope springs eternal. The
    information deciding measurement outcomes must come from somewhere, so
    the environment is a plausible candidate. The success of such a program,
    however, would deny the applicability of the standard interpretation to
    the wavefunction of the universe. MOM assumes from the beginning, that
    this {\em will not work}: The information must come from somewhere other
    than the single state of the standard quantum theory.

\subsection*{Progressive models({\em `Onward and upward!'})}
    Others feel the quantum theory is {\em incomplete}: It needs a broader
    conceptual scheme, or at least new dynamics. This is not a return to
    classical ways, rather speculation on the more {\em radical} features of
    the theory, hoping they will point us in the right direction, towards
    new physics.

{\bf Conscious observer(Wigner):}\cite{Wf,cat}
    In this picture, a measurement outcome results when the mind of a
    conscious observer becomes correlated with the quantum system. It is
    impossible for such a mind to be in a state of fundamental uncertainty
    for long, so instead, an outcome is selected. The conscious observer
    becomes a {\em participator} in the process of reality-building. MOM
    {\em allows} for each state-vector to be associated with an independent
    conscious observer, but this is not necesary, and has no role in the
    dynamics. However, I favor that interpretation. If true, this means that
    consciousness is involved in the resolution of uncertainties above some
    energy scale. At some point, the distinction between the quantum and the
    classical worlds comes about: Which variables resist indeterminancy,
    entanglement with their environment? MOM is a way to explore this issue,
    with or without consciousness.

{\bf Penrose:}
    He presents an alternative to either deterministic or stochastic
    evolution, for state-vector collapse: a {\em non-algorithmic} 
    deterministic process. I.e., although fully determined, quantum
    measurement outcomes can never, even in principle, be predicted. In MOM,
    the chaotic dynamics preclude exact prediction of particular measurement
    outcomes at present, so the distinction is not yet important. For
    Penrose, spacetime variables(e.g. metric) resist quantum indeterminism.
    If he is right, no {\em naive} implementation of `quantum gravity' can
    succeed. Thus, the variables with `observer status' are
    spacetime-structure-related: perhaps at some level, the positions of
    particles, yet at a deeper level, the fabric of spacetime itself. MOM can
    test these ideas by defining the observer cuts appropriately.

{\bf Stochastic models (GRW):} \cite{stoc,stoc2,stoc3,rwalk}
       \cite[Gallis and Fleming]{red}
    In GRW stochastic dynamics modifies the Schroedinger equation,
    effectively driving the system to one of a small set of states. These
    represent the measurement outcomes, and are reached according to BP.
    MOM also assumes some new dynamics, but is deterministic and involves
    multiple quantum states. In the standard measurement case, the following
    picture is useful: The global state undergoes a pseudo-random walk in
    response to the chaotic motion of the local state, which culminates in a
    measurement outcome. This is similar to GRW, but uses chaotic determinism
    in place of randomness.

{\bf Many-minds(Albert):}\cite{Albert}
    This work is an effort to improve upon the many-worlds interpretation by
    incorporating the conscious-observer postulate. I.e., there {\em are}
    multiple universes, wherein measurements have achieved different results,
    and the mind of the observer {\em does} gets split at each
    measurement(really!). There is a different mind for each universe, and
    consciousness is the reason for the measurement process. Yet there is no
    collapse. An attempt is made to place consciousness at center stage,
    rather than classical variables, hence this is a progressive view, 
    unlike MW. MOM assumes that it is {\em not enough} to recognize
    observers have a role. One must also introduce {\em specific} new dynamics
    into the theory to go beyond orthodoxy in a useful way.

\section{Quantum Measurement Categories}
     Here the various schemes discussed above are categorized using some
     obvious criteria, and the results are shown in Figure~\ref{qmschemes}.
     But first we reject the {\bf bare theory without collapse}:
     This is the null interpretation, without deviation from SE. It cannot
     be considered an adequate model of reality, unless measurements
     follow strictly from SE. But this is not so, hence some additional
     structure is needed. Although it has some adherents(also see modal
     interpretation), it leaves little room for discussion.
\subsection*{Observables}
    Are there any classical observables in the theory? These are functions of
    the classical dynamical variables with well-defined values at some or all
    times, when interpreted as quantum operators. If so, are they fixed or do
    they depend on the state of the system or its interactions?

{\bf Space-time(position, gravity):} ({\em GRW, DB, Penrose})
    The observables are associated with the structure of spacetime. Either
    particles' positions (GRW, DB), or the curvature of space (Penrose)
    receive special status: They have well-defined values or resist
    uncertainty.

{\bf Knowledge/mind:} ({\em Wigner, Albert})
    There exists an agent external to the ordinary degrees of freedom of
    the quantum state: the conscious mind. This affects the evolution of
    the system by choosing from the various possible measurement outcomes
    (Wigner). Or it splits into many minds whenever a measurement takes
    place, giving reality to all possibilities (Albert).

{\bf Defined by classical apparatus and interactions:} ({\em Bohr, stat, MW})
    According to Bohr and most physicists, the observable is determined by
    the arrangement of the classical apparatus and its interactions with the
    microsystem. This at once opens possibilities and constrains them: Any
    operator may be an observable, but only if a suitable experimental
    arrangement can be constructed.

{\bf Defined by the quantum state:} ({\em Dec, mod})
    These models rely on the SE evolution and analysis of the state to
    determine observables. Pointer-basis states or decohering histories are
    assigned probabilities which are interpreted classically.

{\bf Defined by multiple states and interactions:} ({\em MOM})
    The observables are defined by the choice of cuts and the interactions
    across them. Thus no specific observable is selected a priori.

\subsection*{Number of Outcomes}
    Does each measurement have a unique outcome, or does anything that
    could happen actually happen, in a set of `parallel universes'?

{\bf One, determined causally:} ({\em DB, MOM})
    In DB a single choice is made for the initial positions of every 
    particle, which determines the outcome of every subsequent measurement.
    In MOM there is also a set of initial data which determines the future
    evolution, including measurement outcomes, but this takes the form of
    quantum states, not classical variables.

{\bf One, non-algorithmic:} ({\em Penrose})
    He proposes that the outcome of a given measurement is determined,
    but in a non-algorithmic way. Hence prediction is impossible in
    principle. The difference between this and random choice may be only of
    philosophical interest.

{\bf One, random:} ({\em Bohr, stat, Wigner, GRW, Dec})
    There is an outcome to each measurement, but it occurs through a
    fundamentally random process, whether for a single event or the history
    of the universe. The reality of the universe, within a quantum
    description, is explained by appeal to mysterious influences beyond
    scientific investigation.

{\bf All possible:} ({\em MW, Albert})
    We give up the notion that there is a single outcome to a given
    measurement: Instead, all possible outcomes actually happen. Does mind
    only follow one path, leaving the rest barren of consciousness, or does
    it split up among the paths, so they all acquire reality? The possible
    existence of parallel universes forces us to consider the role of mind
    in quantum theory.

\subsection*{Process}
    This is usually ignored by most interpreters of quantum theory,
    but otherwise there are only two options.

{\bf Random:} ({\em GRW})
    They use a random Brownian-like motion to nudge the state into one or
    another measurement outcome. It works by replacing a large random effect
    by the concatenation of many small random effects. Should one go further,
    and explain the small random effects by introducing yet smaller random
    effects? I don't think this approach is promising. Rather, one should
    recognize randomness in a physical theory as an indication of a
    {\em temporary} lack of understanding of the processes involved.
    We should not solve one mystery with another.

{\bf Deterministic:} ({\em DB, MOM})
    The basic DB model is entirely deterministic for both the observable
    and the quantum state, but does not allow easy prediction of future
    measurement outcomes. MOM explicitly includes chaotic deterministic
    evolution, so predictions are nearly impossible. However, both models
    offer at least some possibility of {\em prediction} and {\em control}
    of quantum events.

\begin{figure}[!htb]
\setlength{\unitlength}{1cm}
\begin{picture}(16,11)(-11,-4)
  \put(-6,-4){\line(0,1){8}}
  \put(-3,-4){\line(0,1){1.5}}
  \put(-3,-1.5){\line(0,1){2}}
  \put(-3,1.5){\line(0,1){1}}
  \put(-3,3.5){\line(0,1){.5}}
  \multiput(0,-4)(3,0){3}{\line(0,1){8}}
  \put(-6,0){\line(1,0){.75}}
  \put(-3.75,0){\line(1,0){9.75}}
  \multiput(-6,-4)(0,8){2}{\line(1,0){12}}
  \put(-7,2){\makebox(0,0){1}}
  \put(-7,-2){\makebox(0,0){\shortstack{All\\possible}}}
  \put(-8,-2.5){\line(0,1){5}}
  \put(-9,0)
      {\makebox(0,0){\shortstack{{\bf Number}\\{\bf of}\\{\bf outcomes}}}}
  \put(-4.5,4.5){\makebox(0,0){State}}
  \put(-1.5,4.5){\makebox(0,0){Interactions}}
  \put(1.5,4.5){\makebox(0,0){Space}}
  \put(4.5,4.5){\makebox(0,0){Mind}}
  \put(-5,5){\line(1,0){4}}
  \put(1,5){\line(1,0){4}}
  \put(-3,5.5){\makebox(0,0){{\em Generic}}}
  \put(3,5.5){\makebox(0,0){{\em Specific}}}
  \put(-4.5,6){\line(1,0){9}}
  \put(0,6.5){\makebox(0,0){{\bf Observer Prescription}}}
  \put(-3,3){\makebox(0,0){BOHR}}
  \put(-3,1){\makebox(0,0){MOM}}
  \put(-4.5,0){\makebox(0,0){DEC}}
  \put(-3,-2){\makebox(0,0){MW}}
  \put(1.5,3){\makebox(0,0){PENROSE}}
  \put(1.5,1){\makebox(0,0){DB}}
  \put(4.5,2){\makebox(0,0){WIGNER}}
  \put(1.5,-2){\makebox(0,0){GRW}}
  \put(4.5,-2){\makebox(0,0){ALBERT}}
\end{picture}
  \caption[Quantum measurement schemes]
          {Quantum measurement classification, indicating the choices made
           by different schemes. In this analysis, MOM is closer in substance
           to Bohr than any of the others, notwithstanding the philosophical
           differences between the two. \protect}
  \label{qmschemes}
\end{figure}

\section{Outlook}
    Although MOM has been restricted above to simple applications, it is a
    general formalism. However, it is somewhat {\em ad hoc} and hence not yet
    a complete {\em quantum theory of observers}. It seems likely that a more
    fundamental quantum-type theory of nature underlies MOM. Let us consider
    the parameters and other arbitrary aspects of MOM, and how they might be
    constrained.
\begin{description}
  \item[Cyclic(discrete) time $\Delta t$:]
      MOM assumes discrete-time evolution of nature from the start.
      The underlying theory could either provide a continuum description at
      smaller time scales, or prove the impossibility of using continuous
      time, given certain constraints on the evolution. The actual value of
      $\Delta t$ might be found to be the Planck time, or a different time
      scale, or even observer-dependent. Within MOM it remains an arbitrary
      constant, assumed for now universal.
  \item[Decoherence map:]
      The decoherence map itself is somewhat mysterious. It is introduced
      to make the dynamics chaotic, which is crucial to achieve the desired
      results. In a better theory of the time cycle, this could be a kind
      of recurrence map. Perhaps the local state is well defined only at
      specific times, whereas usually it is only potential, inherent in a
      larger structure. A more complicated object would be sometimes
      a member of one Hilbert space or another, but most times something
      more rich in information than any state. This could be a new kind of
      `super-state', or a general density matrix on the global Hilbert space.  
  \item[Universal interaction of observers:]
      This attraction is also new, used to enable the different observers to
      agree on measurement outcomes. Because it involves pairs of states,
      possibly in different Hilbert spaces, the need for a `super-state'
      formulation is even greater here. If two objects interact in some way,
      then they must somehow be {\em very} similar, probably just different
      individuals of the same `species'. The interaction strengths $\beta$
      and $\gamma$ are arbitrary, but there is indication that they depend on
      the dimensionality of the state spaces. The strength should be near
      maximum($\gamma \approx 1$) for a local state in a large environment,
      but very small($\beta \ll 1$) for the global state when the local
      degrees of freedom are few. It may be possible to constrain $\beta$
      from below, if collapse of the wavefunction is real.
  \item[Number, placement and dynamics of cuts:]
      For measurements to occur, the state spaces and the Hamiltonian
      must be right. Interactions must operate across an `active' cut, and
      the decoherence time must be much shorter than $\Delta t$.
      But given a reasonably-sized universal set $\cal U$ of classical degrees
      of freedom, the set of all possible cuts is very large: the power set
      of $\cal U$. Conceivably, every one of the associated Hilbert spaces
      could be populated by any number of quantum states. If each cut
      has even {\em one} state, that is a lot of states. Do we live in such
      a universe? If not, can the active ones `move', their domains 
      changing with time according to some rule? For now, MOM simply assumes
      the active cuts are few in number, their domains fixed.
  \item[Wavefunction of the universe:]
      Whether or not this exists is relevant to quantum cosmology.
      The number of observers/participators/state-vectors could be very large. 
      It is conceivable that there is no `wavefunction of the universe',
      but that instead there is an effective collective state(density matrix)
      with contributions from all observers. Would the underlying theory 
      {\em require} the existence of the wavefunction of the universe, 
      associated with an active cut whose domain contains all other domains?
      So far, MOM assumes a global state exists.
  \item[Multiple states in same Hilbert space:]
      If there are many state vectors in the same space, would this lead
      to different measurement behavior? Their existence could make it seem
      as though a quantum field were present: They would act like a set of
      identical particles. If the underlying theory prescribed which
      cuts are `multiply activated', then it would predict what are the
      `fundamental quantum fields.'
\end{description}

      Let us now consider what kind of measurement result would indicate
      the need for a quantum theory of observers: a detailed account of
      how measurements actually occur.
      All measurements involving micro-systems have so far been of the 
      {\em simple} type. The system of interest, including the micro-system and
      some apparatus, is isolated as well as possible, then it is allowed
      to evolve for some time, and finally a standard measurement is carried
      out on it, using additional apparatus. The initial state of the system
      may be exactly known, or a statistical mixture may be used to describe
      it, but once this initial information is defined, it evolves strictly
      by SE. The results of the measurement, at the end of the isolated
      evolution, can be checked against the predictions of SE and BP.
      They have always turned out to agree, within statistical uncertainty.
      This includes all normal measurements, as well as those checking
      `strange' quantum effects such as the Aharonov-Bohm effect, or EPR-type
      correlations. If experimental results are {\em incompatible} with BP,
      then there are three possibilities:
      a) The theoretical description of the system and its evolution is
      inadequate, because important interactions have been ignored, or
      relevant degrees of freedom left out; 
      b) the predictions of the standard quantum theory, namely BP, 
      are being violated in a fundamental way; or \\
      c) in between the preparation phase, which defined the initial state,
      and the final measurement phase, there was an unexpected,
      spontaneously-occurring measurement of some kind.
      A type (c) result will be called a {\em compound} measurement, because it
      involves additional measurement processes between the preparation phase
      and the final state. Note that this is distinct from the successive
      carrying-out of more than one simple measurement.
      So far, it has always been possible to trace the problem to option (a).
      Thus, there is no evidence for either fundamental deviations from BP,
      or spontaneously-occurring measurements:
      All measurements are {\em caused}
      by explicit actions of the experimenter, and the results {\em always}
      agree with BP. All well-formulated models or theories about the
      measurement process, such as collapse models like MOM, must wait for
      such new experimental results, before they can be judged. These new
      experiments must form the {\em foundation} for the {\em possibility} of 
      studying the measurement process itself. Perhaps some compound
      measurements have already been carried out. These involve
      `intermittent' or `continuous' measurements, with Hamiltonians which
      are not of the pointer-basis type, but contain kinetic as well as
      decohering terms.

\end{document}